\documentclass[acmsmall]{acmart}

\usepackage{tikz}
\usepackage{amsmath}
\usepackage{filecontents}
\usepackage[flushleft]{threeparttable}
\usepackage{amsmath,amsfonts}

\usepackage{graphicx}
\usepackage{textcomp}

 \usepackage{url}

\usepackage{lipsum,tabularx}

\usepackage{multicol}
\usepackage{multirow}

\usepackage{colortbl}
\usepackage{booktabs}
\usepackage{setspace}

\hypersetup{
  linkcolor={blue!70!black},
  citecolor={red!70!black},
  urlcolor={blue!70!black}
}
\def\Snospace~{\S{}}

\usepackage[T1]{fontenc}

\usepackage{algorithm}
\usepackage{algorithmicx}
\usepackage[noend]{algpseudocode}
\usepackage{balance}

\usepackage{bm}
\usepackage{fp}
\usepackage{siunitx}
\usepackage{amsthm}

\usepackage[labelfont=bf,font=small,skip=5pt]{caption}
\usepackage{subcaption}

\usepackage{comment}

\usepackage{fancyhdr}
\usepackage{tikz}
\newcommand*\WC[1]{%
  \begin{tikzpicture}[baseline=(C.base)]
    \node[draw,circle,inner sep=0.2pt](C) {#1};
  \end{tikzpicture}%
}

\usepackage{xspace}

\newcommand{\boxbeg}{
  \vspace{2px}
  \noindent\begin{tabular}{|l|}\hline
    \begin{minipage}{3.2in}
      \vspace{2px}
      \noindent
      }

      \newcommand{\boxend}{
      \vspace{2px}
    \end{minipage} \\ \hline
  \end{tabular}
  \vspace{-10pt}
}

\usepackage{algpseudocode}
\usepackage[htt]{hyphenat}

\usepackage{wrapfig}

\usepackage[scaled=.7]{beramono}
\usepackage{pifont}

\newcommand{\upd}[2]{#2}
\newcommand{\updmj}[2]{#2}
\newcommand{\eg}[0]{e.g.}
\newcommand{\ie}[0]{i.e.}

\AtBeginDocument{%
  \providecommand\BibTeX{{%
    \normalfont B\kern-0.5em{\scshape i\kern-0.25em b}\kern-0.8em\TeX}}}

\setcopyright{acmlicensed}
\acmJournal{TOSEM}
\acmYear{2025} \acmVolume{1} \acmNumber{1} \acmArticle{1} \acmMonth{1}\acmDOI{10.1145/3773993}

\begin{document}

\newcommand{\sys}{\textsc{Citadel}\xspace} 
\title[\sys: Context Similarity Based Deep Learning Framework Bug Finding]{\sys: Context Similarity Based Deep Learning Framework Bug Finding}

\author{Xiaoyu Zhang}
\orcid{0000-0001-7010-6749}
\affiliation{%
  \institution{Xi'an Jiaotong University}
  \city{Xi'an}
  \country{China}
}
\email{zxy0927@stu.xjtu.edu.cn}

\author{Juan Zhai}
\orcid{0000-0001-5017-8016}
\affiliation{%
  \institution{University of Massachusetts, Amherst}
  \country{United States}}
\email{juanzhai@umass.edu}

\author{Shiqing Ma}
\orcid{0000-0003-1551-8948}
\affiliation{%
  \institution{University of Massachusetts, Amherst}
  \country{United States}}
\email{shiqingma@umass.edu}

\author{Shiwei Wang}
\orcid{0000-0002-2777-7392}
\affiliation{%
  \institution{Xi'an Jiaotong University}
  \country{China}}
\email{shiwei.wang@stu.xjtu.edu.cn}

\author{Chao Shen}
\orcid{0000-0002-6959-0569}
\authornote{Chao Shen is the corresponding author.}
\affiliation{%
  \institution{Xi'an Jiaotong University}
  \country{China}}
\email{chaoshen@xjtu.edu.cn}

\renewcommand{\shortauthors}{Zhang and Zhai, et al.}

\begin{abstract}

With the application of deep learning technology, tools of DL framework testing are in high demand.
Existing DL framework testing tools have limited coverage of bug types.
For example, they lack the capability of effectively finding performance bugs, which are critical for DL models regarding performance, economics, and the environment.
Moreover, existing tools are inefficient, generating hundreds of test cases with few trigger bugs.
In this paper, we propose \sys, a method that accelerates bug finding in terms of efficiency and effectiveness.
We observe that many DL framework bugs are similar due to the similarity of operators and algorithms belonging to the same family.
Orthogonal to existing bug-finding tools, \sys aims to find new bugs that are similar to reported ones that have known test oracles.
\sys defines \textit{context similarity} to measure the similarity of DL framework API pairs and automatically generates test cases with oracles for APIs that are similar to the problematic APIs in existing bug reports.
\upd{Response to R3Q1: }{
\sys effectively detects 58 and 66 API bugs on PyTorch and TensorFlow (excluding those rejected by developers or duplicates of prior reports), many of which, \eg, 13 performance bugs, cannot be detected by existing tools.
}
Moreover, 35.40\% of test cases generated by \sys can trigger bugs significantly transcending the state-of-the-art method (3.90\%).
\end{abstract}

\ccsdesc[500]{Software and its engineering~Software libraries and repositories}

\keywords{Deep Learning Testing, Deep Learning Library, Software Testing}

\maketitle

\section{Introduction}\label{sec:intro}

With the development of Deep Learning~(DL) techniques, DL-powered systems are playing an increasingly significant role in software development. 
For example, Microsoft has developed a new search engine powered by DL techniques to enhance the search results~\cite{microsoftnewbing}.
Moreover, the global AI software market is forecasted to increase from \$257 billion in 2025 to \$1,459 billion by 2034~\cite{marketsandmarkets}.
As the backbone of DL-powered systems, DL frameworks (\eg, TensorFlow and PyTorch) empower developers by offering API functions to create, train, optimize, and deploy DL-powered systems.
These frameworks support diverse domains, providing societal benefits in areas like image recognition~\cite{he2016deep}, self-driving~\cite{grigorescu2020survey}, and natural language processing~\cite{li2018deep}.  
Similar to traditional software systems, DL frameworks can also have bugs, which can lead to erroneous outputs, increased system overhead, and even crashes for DL-powered systems, thereby jeopardizing user property and personal safety~\cite{teslacrash}, 
and contributing to energy inefficiency and environmental issues~\cite{nistor2015caramel,jin2012understanding,lacoste2019quantifying}.
Consequently, there is a pressing need for tools capable of identifying bugs in DL frameworks.

There are two primary approaches to testing DL frameworks: model-level testing~\cite{pham2019cradle,guo2020audee,wang2020deep,gu2022muffin} and API-level testing~\cite{wei2022free,deng2022fuzzing}.
Model-level testing mutates existing DL models to generate more diverse DL models and employs differential testing methods to compare model execution results across different frameworks for bug detection.
In contrast, API-level testing approaches generate test code directly for DL framework API functions, exposing bugs through fuzzing techniques.
For example, DocTer~\cite{xie2022docter} conducts fuzzing for DL frameworks by extracting input constraints from API documentation and using these constraints to guide test case generation.
DeepREL~\cite{deng2022fuzzing} identifies relational API functions of DL frameworks and `borrows' test inputs from invoked API functions to test other relational API functions.

Despite these advancements, existing DL testing tools have notable limitations.
Firstly, existing testing tools have limited coverage of bug types.
For example, they can hardly detect performance bugs that can significantly impact DL model training and inference speed and degrade responsiveness, resulting in energy waste and environmental concerns~\cite{nistor2015caramel,jin2012understanding,lacoste2019quantifying,bryant2003computer}, especially for large DL models like GPT-3~\cite{patterson2021carbon}.
The performance bug shown in~\autoref{fig:moti} causes the time overhead to increase to 2.33 times its original value and a substantial carbon footprint.
However, current testing methods cannot detect such performance bugs in DL API functions.
Secondly, existing bug-finding tools exhibit inefficiencies in generating test cases that trigger bugs.
These tools often leverage random walks or heuristic algorithms to generate test cases. 
However, due to the huge search space of API arguments and inputs, such approaches often generate numerous test cases, but only a small fraction of them trigger actual bugs.
For example,
DeepREL generates an excess of 330,000 test cases, yet only 1.23\% of them have the potential to trigger bugs. Requiring hundreds of test cases to detect a single bug makes the current DL framework testing tools very inefficient in detecting bugs.

To devise an efficient test method capable of identifying various types of bugs, we thoroughly analyze the API functions of PyTorch and TensorFlow and study their reported issues on GitHub.
The API functions of these frameworks naturally fall into distinct groups, where API functions within each group execute similar operators and algorithms, exhibiting a tendency for similar bugs.
Considering the convolutional operators \texttt{torch.nn.Conv1d}, \texttt{torch.nn.Conv2d}, and \texttt{torch.nn.Conv3d}, each is designed for inputs of different dimensions.
Despite their differences, these operators share commonalities such as call lists (\eg, \texttt{aten::convolution} in the source code) and the use of the \texttt{cudaLaunchKernel} function for GPU computations.
Notably, reported issues~\cite{torch_issue1,torch_issue2} emphasize that when a bug arises in one convolution operator, other operators within the same group are prone to similar bugs.

Building upon this observation, we propose \sys, a tool that accelerates the finding of bugs in terms of efficiency and effectiveness.
\upd{Response to R3Q3: }{
Orthogonal to existing tools that explore new anomalous behaviors and report bugs, \sys aims to uncover new bugs that are similar to reported ones that have known test oracles, regardless of bug types.
}
It leverages reported bugs on one API function to create test cases for its analogous API functions, effectively addressing the aforementioned limitations observed in existing work, and can easily co-work with other testing methods to accelerate bug discovery.
Compared with existing tools, which can only detect status bugs and value bugs, \sys has better bug type coverage and effectiveness.
It has the capability to detect bugs regardless of their types, such as performance bugs caused by errors in the underlying implementation or optimization, including unexpected time or memory overhead.
Moreover, \sys is more effective and efficient in test case generation.
It adopts the code that has triggered a bug on a problematic API function to create test cases for its analogous API functions.
Essentially, it leverages prior knowledge rather than heuristics to explore potential API bugs in the new context, significantly improving the chances of finding bugs.
To be specific, we first collect existing bug reports and identify problematic APIs.
\sys then utilizes both static and dynamic analyses on the DL framework source code and unit test cases for the identification of analogous API functions.
In this process, it extracts context information (\eg, APIs' call stacks) to gauge the similarity between API functions, a concept referred to as \emph{context similarity} in \sys.
For a collected problematic API, \sys modifies the bug-triggering code from its bug report to generate new test cases for its analogous API functions.
Throughout this process, \sys addresses two potential differences between the API functions: differences in arguments and dimensions, if they exist.
Finally, \sys executes the generated test cases, employing the buggy behavior of the problematic API function as a test oracle to effectively identify potential bugs in the target API function.

Our evaluation demonstrates that \sys successfully identified a total of 77 API bugs in PyTorch and 74 in TensorFlow, including 58 and 66 previously unreported bugs, of which 36 and 56 have been confirmed. Additionally, 49 of these bugs are detected by analogous API pairs that existing approaches do not cover. 
Furthermore, a noteworthy 35.40\% of the test cases synthesized by \sys expose bugs, significantly surpassing the 0.74\%, 1.23\%, and 3.90\% bug-triggering capacity exhibited by the test cases generated by DocTer, DeepREL, and TitanFuzz, respectively.

Our contributions are:
\begin{itemize}
    \item We propose \textit{context similarity} as a measurement for functional similarity among DL framework API functions.
    \item We develop a novel test case generation method for DL frameworks that leverages the knowledge from confirmed API bugs to synthesize new test cases and uncover bugs in analogous API functions, regardless of bug types.
    \item We develop a prototype \sys based on the proposed idea.
    \upd{Response to R3Q1: }{The experimental results on PyTorch and TensorFlow show that \sys detects 58 and 66 previously unknown API bugs, respectively, among which 36 and 56 have been confirmed or fixed by developers after reporting.}
    35.40\% of test cases generated by \sys can be used to trigger bugs.
\end{itemize}

\section{Background}\label{sec:bg}


\subsection{DL Framework API Functions and Models}\label{sec:bgmodel}

\noindent
{\bf DL Framework APIs.}
Like traditional software programs, DL frameworks use various API functions to call source code functions and perform operations.
Taking PyTorch~\cite{paszke2019pytorch} as an example, its API functions include performing basic matrix operations (\eg, \texttt{torch.mul} for multiply operation), calculating loss functions (\eg, \texttt{torch.nn.MSELoss} for measuring mean squared error), and building model layers (\eg, \texttt{torch.nn.Conv2d} for convolution layers).
When calling an API function, users first need to assign values for its required and optional arguments, where the values of required arguments are mandatory to provide, and the optional arguments have their default values in APIs.
Then, the API function runs the underlying source code that performs corresponding calculations and operations on the hardware (\eg, CPU and GPU) and obtains tensors, Boolean values, etc. as the result.

\noindent
{\bf DL Models.}
A DL model is a parameterized function \(F_\theta: X \mapsto Y\), where \(x\in X\) is an \(m-\)dimensional input and \(y\in Y\) is the corresponding output label.
Typically, a DL model is composed of several connected layers, and an $n$-layered model can be represented as \(F_\theta =  l_1 \circ l_2
\circ \cdot\cdot\cdot \circ l_n\), where \(l\) represents a layer and \(\theta\) is the model weight.
Each layer \(l_i\) in the model can be constructed by several DL framework API functions.
Therefore, a DL model can also be represented as a directed acyclic graph (DAG) in that API functions are nodes, and the returned values of API (\eg, tensors) are edges.
Running and training a DL model \(F\) on the input-output pairs (\(x_i, y_i\)) is essentially calling a series of API functions and passing their outputs based on the topological sorting of its computation graph~\cite{deng2022fuzzing}.

\subsection{DL Framework Testing}\label{sec:dltest}

DL framework testing methods construct test cases (\eg, models) to explore abnormal behaviors of DL frameworks and discover bugs
Depending on the generated test cases, existing DL framework testing methods can be mainly divided into model-level testing and API-level testing~\cite{wei2022free,deng2022fuzzing}.

\noindent
\textbf{Model-level testing.}
These testing methods usually build a large number of models and apply mutation strategies on models to explore the potential bugs of the APIs and layers in the model.
To construct test oracles, prior work performs differential testing by building and testing the same model on multiple DL frameworks~\cite{guo2020audee,pham2019cradle,gu2022muffin,wang2020deep}.
CRADLE is one of the first to use this method to test DL framework bugs.
Based on Keras~\cite{ketkar2017introduction}, which can build and train models on different DL frameworks as backends, it conducts differential testing on three frameworks (\ie, TensorFlow, CNTK, and Theano) and finds 12 bugs.
Additionally, Muffin~\cite{gu2022muffin} creatively designs the data tracking method to apply differential testing on the training phase of models and finally discovers 39 new bugs.
Although the model-level methods obtain outstanding test results, they still have great limitations in applications.
Due to limitations imposed by the test model, these methods typically support only a limited subset of API functions related to the models. For instance, existing research~\cite{wei2022free} reports that LEMON covers only 35 TensorFlow APIs. 
Furthermore, since the test oracle relies on the implementations of multiple frameworks, inconsistencies detected during testing are often difficult to verify as real bugs, which hampers the effectiveness of the bug detection process~\cite{guo2020audee, pham2019cradle}.

\noindent
\textbf{API-level testing.}
Different from the model-level methods, the API-level framework testing methods do not depend on the implementations of multiple frameworks and have the capability to test abnormal behaviors of more API functions.
API-level testing usually extracts API constraints of inputs and arguments based on the documentation or test code and generates test cases based on the fuzzing technique~\cite{xie2022docter,zhang2021predoo,deng2022fuzzing,wei2022free,christou2023ivysyn,yang2023fuzzing,deng2024large}.
DocTer~\cite{xie2022docter} analyzes the API document syntax and extracts input constraints.
It can generate test cases for three different DL frameworks and find 94 bugs on these frameworks.
EAGLE~\cite{wang2022eagle} proposes that some APIs have functional equivalence.
It designs 16 new DL equivalence rules and detects 25 inconsistencies and bugs on TensorFlow and PyTorch.
In addition, DeepREL~\cite{deng2022fuzzing} designs two elaborated equivalence relations and matches API pairs based on these equivalence relations.
It considers the output values and status of APIs in a pair as test oracles for each other and detects both crash and inconsistency bugs for over 1,000 PyTorch API functions.
TitanFuzz~\cite{deng2023large} and \(\nabla\)Fuzz~\cite{yang2023fuzzing} leverage large language models (LLMs) and automatic differentiation to generate test code and implement API-level fuzzing to detect numerical inconsistencies and crashes in DL frameworks.
\updmj{Response to R2Q2: }{
Note that localizing the root cause of APIs' abnormal behaviors is a challenging task, often requiring significant time and effort.
Existing methods~\cite{xie2022docter,deng2022fuzzing} typically count the number of abnormal behaviors in different APIs (i.e., API bug in this paper) without distinguishing whether they share the same root cause or implementation errors.
Following the prior work, \sys leverages existing bug reports to effectively find and report bugs on analogous API functions without being limited by bug types.
}

\noindent
\textbf{DL framework bugs.}
DL framework bugs can be mainly divided into three types through symptoms, \ie, status, value, and performance bugs~\cite{chen2022toward}.
Status bugs affect the execution status of the DL API and model, including crashes, segmentation faults, exceptions, etc.
Value bugs that are caused by numerical errors in the computation of DL operators include inconsistent outputs and NaN (Not A Number) outputs.
Existing framework testing tools focus on the above two types of bugs~\cite{guo2020audee,deng2022fuzzing,xie2022docter}.
Performance bugs refer to those caused by errors in the underlying implementation or optimization, including unexpected time or memory overhead.

\subsection{Code Similarity Measurement}
\label{sec:scmeasure}

Code similarity measurement aims to evaluate the similarity of multiple code blocks and find potential code clones, plagiarism, and refactoring.
Existing static approaches proposed methods based on the metrics, texts, and tokens~\cite{faidhi1987empirical,luo2014semantics,misu2017interface,wang2023ccstokener}.
Researchers also measure code similarity based on Abstract Syntax Trees (ASTs) and graphs (\eg, control flow graphs (CFG))~\cite{zhang2019novel,chae2013software}.
In addition, some research proposes the functional similarity between programs from the perspectives of input and output and function calls~\cite{su2016identifying,mcmillan2012detecting}, etc.
However, these methods and tools are usually designed for code snippets in one single programming language, but DL framework API functions execute on both Python and C++ source code and involve various wrappers, which poses a challenge in evaluating the similarity between these API functions.
Inspired by existing approaches, \sys defines and calculates the context similarity to match and test DL framework API functions.
\sys calculates the similarity of source code blocks from the perspectives of inputs, outputs, and functions.
Then it combines the above results with the called functions and traces of different DL framework APIs to match APIs that have a similar functionality and execution context.

\section{Motivation}
\label{sec:moti-new}

State-of-the-art DL framework testing tools~\cite{xie2022docter,deng2022fuzzing,gu2022muffin} have two major limitations.

\noindent
\(\bullet\)
\upd{Response to R2Q2: }{
{\bf Existing approaches have limited coverage on bug types.
They focus on status and value bugs, lacking the capability of effectively detecting others, \eg, performance bugs.}
Existing methods typically utilize the differential testing techniques to construct pseudo test oracles for bug detection.
These approaches compare the output results of the same or equivalent API functions on different frameworks/devices to identify status bugs (\eg, crashes) and value bugs (\eg, NaN outputs)~\cite{wang2022eagle,gu2022muffin,deng2022fuzzing,guo2020audee,pham2019cradle,deng2023large}.
However, they are generally unable to construct test oracles to detect performance bugs due to the difficulty of obtaining test oracles.
In addition, although some metamorphic testing methods have successfully identified several performance bugs~\cite{wei2022free}, their effectiveness is constrained by the manually designed metamorphic relations.
As a result, they can only test the specific API behaviors (e.g., those related to tensor types) and are unable to identify broader categories of performance bugs (e.g., the \texttt{LazyConvTranspose2d} bug in~\autoref{fig:moti}).
}

\noindent
\(\bullet\)
{\bf Existing methods need to generate numerous test cases to trigger a bug, resulting in inefficient testing.}
To uncover bugs within DL frameworks, existing work usually leverages random walks or heuristic algorithms to generate test cases and models, exploring potential API behaviors~\cite{xie2022docter,deng2022fuzzing,guo2020audee,gu2022muffin}.
On the one hand, considering the vast search space of the arguments and inputs of API functions, the random method has a low probability of generating a test case that reveals a bug.
On the other hand, the heuristic algorithm (\eg, Genetic Algorithm) typically requires the construction of large populations and multiple generations of mutation to search for bugs, rendering them impractical.
Moreover, whether the evaluation of the heuristic algorithm can effectively guide the testing is questionable.
Consequently, existing work typically needs to generate hundreds of test cases to uncover a bug, resulting in inefficient testing on the DL framework.

\section{Design}\label{sec:design}

\begin{figure}
    \centering     
    \includegraphics[width=0.7\linewidth]{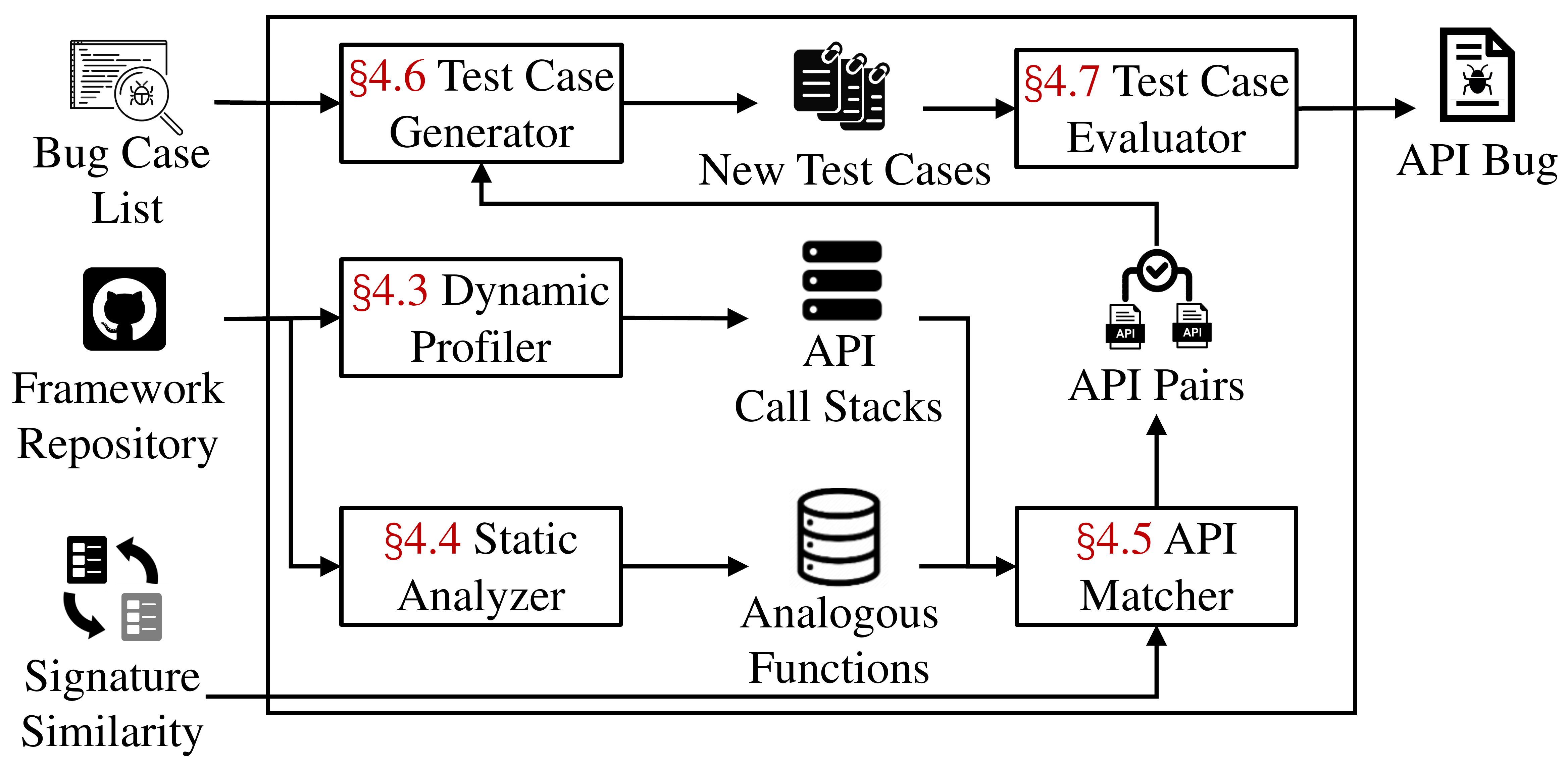}
    \caption{Overarching Design of \sys}\label{fig:overview}
\end{figure}



\updmj{Response to R2Q1: }{
We observe that DL framework API functions naturally fall into groups.
For example, in PyTorch, while APIs like \texttt{Conv1d}, \texttt{Conv2d}, and \texttt{Conv3d} expose parameter signatures designed to accommodate different data dimensions,
they call similar or even the same operators and algorithms during runtime, indicating that their functionality and implementation are highly similar.
Existing research has confirmed that this phenomenon of code cloning is prevalent in DL frameworks~\cite{assi2024unraveling}.
Code cloning not merely significantly increases software maintenance costs but also accelerates the propagation of defects, that is, implementation errors in one API are highly likely to exist in other analogous API functions within the same group~\cite{islam2016bug}.
To verify this hypothesis, we sample and manually analyze over 300 real bug reports and corresponding patches from DL frameworks and find that API functions within the same group are highly susceptible to similar or even the same implementation errors, leading to a series of bugs across multiple APIs~\cite{torch_issue1,torch_pr1,torch_pr2}.
For example, prior work~\cite{xie2022docter} has detected a series of crash bugs on the \texttt{conv1d}, \texttt{conv2d}, and \texttt{conv3d} APIs of PyTorch, which actually have the same implementation error (i.e., missing checks on the variable \texttt{groups}) and can eventually be fixed by the same patch\footnote{https://github.com/pytorch/pytorch/pull/77919}.
Based on the above findings, this paper proposes CITADEL, which aims to systematically exploit the similarities in functionality and implementation between APIs (i.e., contextual similarity) and combine them with known bug reports to efficiently discover bugs in DL framework APIs.
}

\autoref{fig:overview} shows the overview of \sys.
The inputs to \sys include DL framework repositories, real-world bug cases collected from the framework issues, and the signature similarity between DL API functions~\cite{deng2022fuzzing}.
Specifically, in this paper, \sys is applied to the PyTorch and TensorFlow repositories, two of the most widely used DL frameworks, which have garnered 87K and 189K stars on GitHub, respectively.
Bug cases are collected from the issue lists of these two frameworks, including reproducible buggy code and problematic API functions (\autoref{sec:sampler}).
The workflow of \sys begins by extracting context information from the DL framework's repository, including the API function call stacks and analogous function groups derived from the framework’s source code.
In this process, the dynamic profiler (\autoref{s:dynamic}) generates unit test cases for DL API functions and records their call stacks, capturing the source code functions invoked during execution.
Simultaneously, the static analyzer (\autoref{s:static}) examines the DL framework’s source code and clusters analogous functions based on argument and callee similarity.
\sys proposes \textit{context similarity}, which utilizes API call stacks to measure the similarity between API functions.
Additionally, the analogous source code functions are treated as identical during similarity calculation.
Leveraging the context similarity and the signature similarity~\cite{deng2022fuzzing}, the API matcher (\autoref{sec:apimatcher}) matches analogous API pairs.
Furthermore, the matcher verifies the arguments of each API pair and discards those with unsolvable argument mismatches.
For a problematic API function (\ie, source API) and its analogous API function (\ie, target API), the test case generator utilizes the reproducible buggy code of the source API, which is collected in the bug case list, to synthesize new test cases for the target API (\autoref{sec:generate}).
The test case evaluator then executes new test cases and leverages the buggy behavior exhibited by the source API to identify bugs in the target API, including status, value, and performance bugs (\autoref{sec:evaluator}).
Finally, \sys reports the newly detected API bugs to the user.


\begin{figure}
    \centering     
    \includegraphics[width=0.7\textwidth]{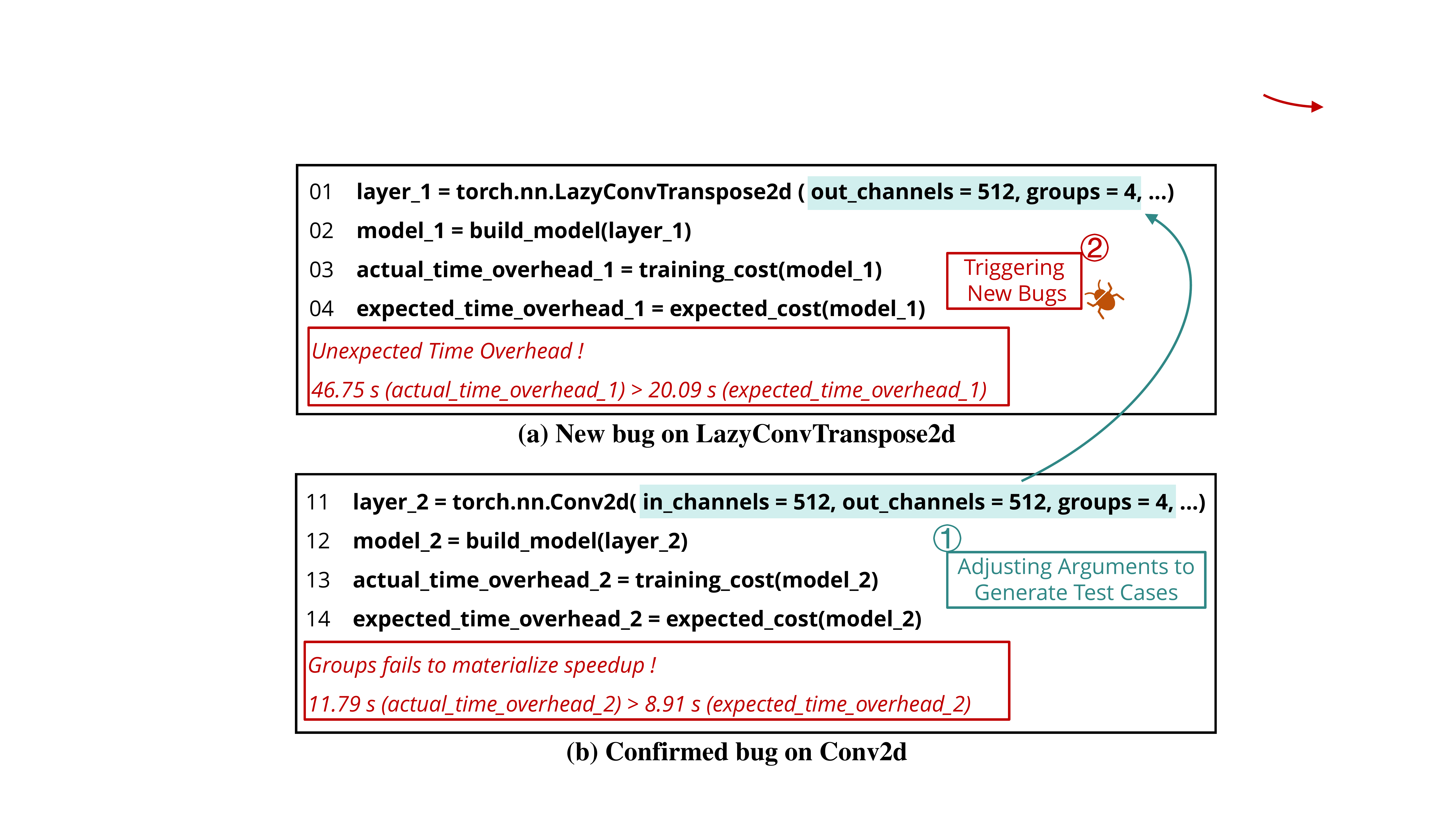}
    \caption{Performance Bug on \texttt{LazyConvTranspose2d}}
    \label{fig:moti}
\end{figure}

\noindent
\textbf{\sys in an example.}
\sys detects a total of 151 API bugs, including 103 status bugs, 35 value bugs, and 13 performance bugs.
Moreover, 35.40\% of test cases generated by \sys can trigger bugs, and this ratio is only 0.74\% and 1.23\% in DocTer, DeepREL, respectively.

Here we provide a real-world performance bug found by \sys as an example (Lines 1-4 in ~\autoref{fig:moti}(a)) to show how it works.
One collected bug on \texttt{Conv2d} (\autoref{fig:moti}(b)) reports that the `groups' argument in this function fails to speed up the training and inference.
Grouped convolution aims to employ multiple kernels and produce multiple channel outputs to increase the network efficiency~\cite{krizhevsky2017imagenet,gong2021deep}.
Therefore, the group convolution is anticipated to bring a lower time overhead compared to executing these convolution layers independently.
The code in the collected bug uses the execution time of independent convolution layers to estimate an upper bound on expected time overhead of group convolution, which is accepted by the developers, and finds that the actual overhead of group convolution (11.79~s) is much greater than the expected upper bound (8.91~s), therefore identified the performance bug on \texttt{Conv2d}.
\sys analyzes the context information (\eg, call stacks) of DL API functions to construct pairs of analogous API functions that share context similarity.
One such pair consists of \texttt{Conv2d} and \texttt{LazyConvTranspose2d}.
Then, \sys generates test cases for \texttt{LazyConvTranspose2d} based on the reproducible code of the problematic API \texttt{Conv2d}.
For each analogous API pair, \sys analyzes the arguments of the two API functions to identify differences and makes adjustments to the buggy code accordingly to construct a test case for the target, ensuring the generated test case is executable.
In this instance, \sys remove `in\_channels' to resolve the difference between APIs' arguments, which is highlighted by green (\WC{1}).
In addition, \sys leverages the method in the source bug report of \texttt{Conv2d} to estimate the expected upper bound on the time overhead of the grouped \texttt{LazyConvTranspose2d} layers using the overhead of independent \texttt{LazyConvTranspose2d} layers.
The new test case reveals that \texttt{LazyConvTranspose2d} with `group' argument also exhibits a higher time overhead than executing these layers individually, which is the same anomalous behavior as the reported bug in \texttt{Conv2d} (\WC{2}).
Specifically, as shown in~\autoref{fig:moti}(a), when we set the `out\_channels' to 512 and `groups' to 4 in \texttt{LazyConvTranspose2d} layer and construct \textit{model\_1} with eight such group convolution layers, the actual time cost of training \textit{model\_1} is 46.75 s, which is markedly higher than the time cost of executing these layers individually (20.09~s).
Finally, \sys discovers a performance bug on the \texttt{LazyConvTranspose2d} API function, which has been confirmed by developers~\cite{perfbug1}.

\subsection{Preparation: Bug cases Collection}
\label{sec:sampler}

We implement a bug case sampler to extract reproducible bug cases from the latest tens of thousands of GitHub bug issues (both open and closed) of the two DL frameworks as the \sys's input, discarding issues that lack reproducible code.
Currently, PyTorch and TensorFlow provide well-structured issue templates for bug reporting~\cite{torch_template,tf_template}.
These templates request minimal and complete code examples to reproduce the bug and the anomalous behaviors.
Taking the \texttt{Conv2d} bug in~\autoref{fig:moti} as an example, its report contains executable code to call the buggy convolution layers (Lines 11-12 in~\autoref{fig:moti}).
Additionally, its code calculates the expected time overhead of the group convolution and contrasts it with the actual time cost to directly demonstrate the buggy behavior (Lines 13-14).


\begin{figure}
    \centering     
    \includegraphics[width=0.75\textwidth]{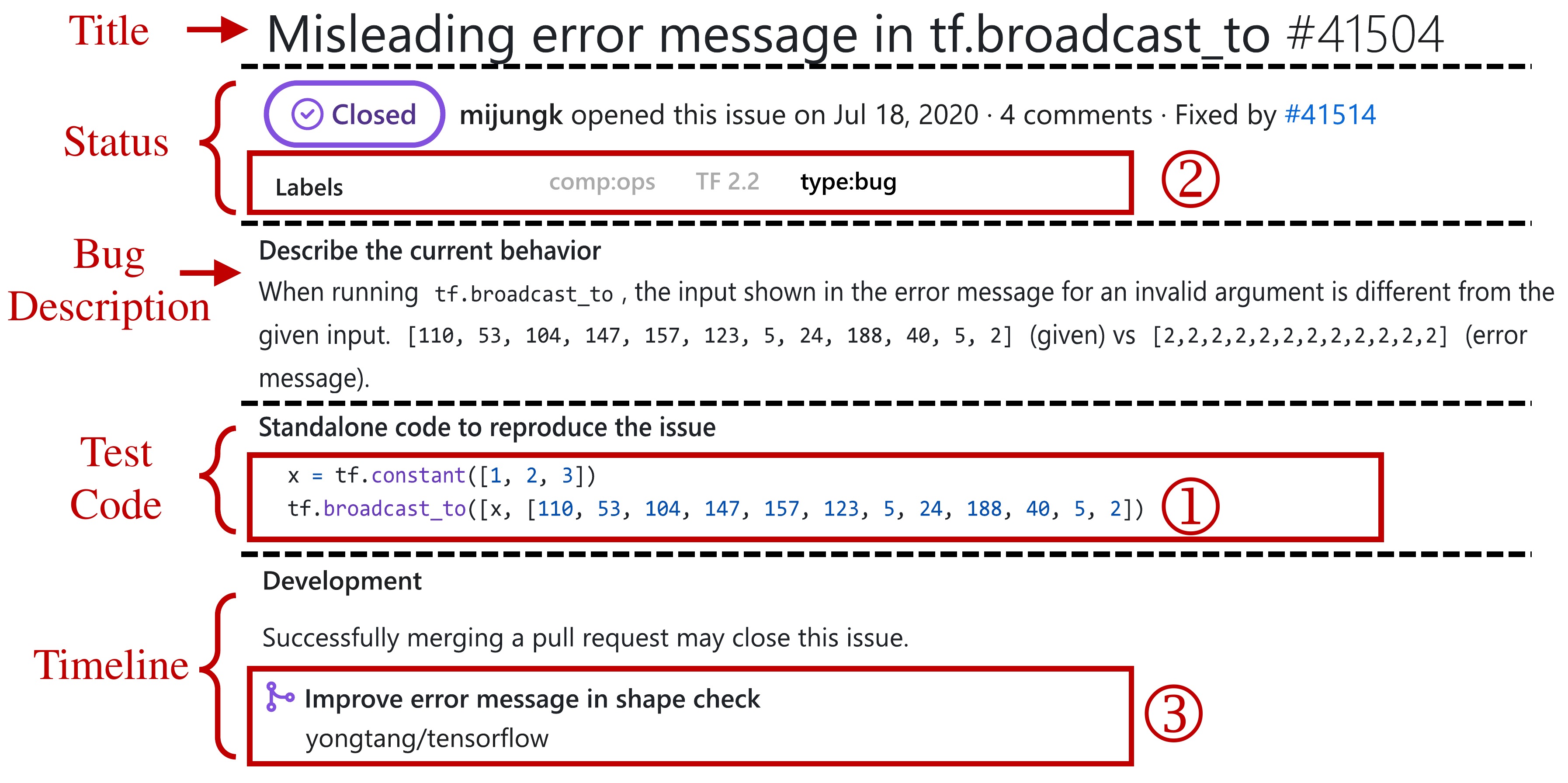}
    \caption{A Demo of Sampling Bug cases}
    \label{fig:sampler}
\end{figure}
\autoref{fig:sampler} shows a real report from the TensorFlow repository, consisting of the title, status, bug description, test code, and a timeline.
To extract the buggy code, the sampler first judges whether one issue includes test code (\WC{1}) and discards those lacking test code.
Then, it checks issue labels (\WC{2}) and discards reports that are not marked as bugs, crashes, etc. by developers.
These issues usually report non-bug problems (\eg, documentation typos) and are not assigned labels by developers.
During this process, the sampler also discards reports related to specific hardware (\eg, M1 chips~\cite{torch_issuem1}).
Limited by the experiment environment, we cannot reproduce these bugs.
Finally, the sampler examines the issue timeline (\WC{3}) and discards issues with fewer than 3 comments, which are usually reports of issues that developers do not care about or intend to work on~\cite{torch_issuecomment}.
It also discards closed issues that lack associated commits or pull requests.
Such issues often arise from users' misconceptions of expected behaviors and are promptly addressed by developers~\cite{torch_expect2}.
After extracting buggy code, the sampler matches the most frequently mentioned API function from the `Title' and `Bug Description' (as shown in~\autoref{fig:sampler}) as a problematic API candidate and verifies it in the corresponding test code.
If the candidate does not appear in the code, the second most mentioned API is selected, and so on.
With such a method, the problematic API in~\autoref{fig:sampler} can be correctly identified as \texttt{tf.broadcast}.
Detailed implementation of the sampler is in our repository~\cite{our_repo}.

The bug cases collected by the sampler will be manually verified to determine whether the reported buggy behavior can be reproduced and to classify their bug types (\ie, status, value, or performance bug).
Specifically, we invite two co-authors in the fields of software engineering and artificial intelligence to review the collected cases and label their bug types based on the taxonomy in~\autoref{sec:dltest}.
In addition, since \sys cannot currently leverage performance bug cases that describe expected overhead in natural language or images~\cite{torch_perfnlp} to generate new test cases, during the manual review, we only retain those with available code for calculating and estimating the expected overhead.
For inconsistent review results, we invite a third co-author to lead the discussion until the review results are recognized by all three.

We acknowledge that, similar to prior approaches, CITADEL also requires a certain amount of manual effort, mainly to verify and ensure the effectiveness of the test cases extracted from bug reports, as described in Section 4.1..

\upd{Response to R2Q3:}{
The manual effort at this preparation stage is mainly to verify and ensure the effectiveness of the test cases extracted from bug reports, which will be used in the following experiments (\autoref{s:effective}).
Note that in some application scenarios, once the effectiveness of the test cases is ensured, the associated manual effort can be significantly reduced or even eliminated.
For example, when \sys is used in conjunction with other fuzzing tools, it can directly leverage the test cases that are generated by other tools and have triggered individual API bugs as input.
\sys can automatically construct new test cases for analogous APIs without requiring additional verification, thereby enabling effective and efficient bug detection.
}

\subsection{Context Similarity}\label{s:context}

Context similarity calculates the similarity between the runtime context of APIs, and the similar contexts intuitively show that the functionality of APIs would be similar.
We observe that on some DL framework APIs (\eg, \texttt{Conv1d}, \texttt{Conv2d}, and \texttt{Conv3d} in PyTorch), although there are differences between their inputs or arguments (\eg, different dimensions), they have similar functionalities and implementations, which can be reflected by context similarity.
Many issues and patches~\cite{torch_issue1,torch_pr1,torch_pr2} in GitHub further reveal that API functions that have similar functionalities are prone to have similar bugs due to one erroneous implementation of an underlying function.
Moreover, the prior work~\cite{wang2022eagle} has proposed that the functional similarity between API functions can guide the construction of equivalence rules in testing.
Therefore, we define and measure the context similarity between API functions to find API functions that have similar functionalities and leverage bugs on one API function to effectively identify potential bugs on its analogous API functions.

Specifically, we measure the context similarity \(Sim_{CTX}(A_S,A_T)\) for any API pair \((A_S, A_T)\) in \sys:
$$
Sim_{CTX}(A_S,A_T) = J(CTX_{S}, CTX_{T}),
$$
where \(CTX_{S}\) and \(CTX_{T}\) represent the context information of \(A_S\) and \(A_T\), which is collected by the static analyzer and dynamic profiler.
\(J\) indicates the metric to calculate the similarity between \(CTX_{S}\) and \(CTX_{T}\).
In this paper, we use Jaccard similarity coefficient~\cite{murphy1996finley} as \(J\) to calculate \(Sim_{CTX}\).
The greater the similarity between the execution contexts of two API functions, the higher the probability that they have similar underlying implementations and perform similar operators, and they are also more susceptible to similar bugs. 

\subsection{Dynamic Profiler}\label{s:dynamic}

The most important context information of APIs is the source code functions they call during execution (\ie, the call stack), which intuitively demonstrates the underlying implementation of the API functions.
To effectively collect such context information, the dynamic profiler executes unit test cases for API functions and records call stacks.
The unit test cases consist of the test cases collected from DL framework repositories and the test cases generated by existing test case generation tools~\cite{xie2022docter,deng2022fuzzing}.
These cases are intended to examine the expected behaviors of APIs during runtime and explore the analogous behaviors and edge cases.
\autoref{fig:dynamic} illustrates how \sys extract context information and matches \texttt{Conv2d} and \texttt{LazyConvTranspose2d} in the~\autoref{fig:moti} as analogous APIs.
The solid box in~\autoref{fig:dynamic} illustrates part of the API call stacks collected by the profiler.
During execution, both APIs call the source code function \texttt{aten::convolution} for performing convolution operations and \texttt{cudaLaunchKernel} related to GPU services.
In addition, \texttt{Conv2d} calls the source code function \texttt{aten::conv2d} related to the optimization of 2D convolution, while \texttt{LazyConvTranspose2d} calls \texttt{aten::conv\_transpose2d} related to transpose convolution.

\subsection{Static Analyzer}\label{s:static}

Existing research~\cite{assi2024unraveling} has revealed that a significant number of code blocks and functions with similar implementations exist within the source code of DL frameworks. 
Source code functions exhibiting similar functionality and implementation patterns may have similar errors, leading to bugs in the DL APIs that invoke them.
The static analyzer is designed to identify and cluster these analogous functions within the source code.
It then incorporates them as part of the context information provided to the API matcher, facilitating the measurement and matching of APIs with similar execution contexts. 
Inspired by the prior work~\cite{alkhalaf2014semantic,su2016identifying,mcmillan2012detecting,novak2019source}, the static analyzer determines whether two functions, \(F_1\) and \(F_2\), have similar implementations by evaluating two key aspects: input and output arguments similarity \(Sim_{io}\) and callees similarity \(Sim_{call}\).

\noindent
{\bf Input and output arguments} play a crucial role in determining functional similarity in code blocks~\cite{alkhalaf2014semantic,su2016identifying}.
For a function \(F_1\) within the source code, the static analyzer captures its input and output arguments and formalizes them as a set \(IO_{F_1}=\{a^1_1, a^1_2, ....,a^1_n\}\), where \(a^1_i\) represents an input or output argument of \(F_1\).
To evaluate the similarity of two sets, we use the Jaccard similarity coefficient~\cite{murphy1996finley}, a widely utilized statistical measure~\cite{svajlenko2017fast, ragkhitwetsagul2018picture}.
The Jaccard similarity coefficient \(J(A, B)\) for two given sets \(A\) and \(B\) is defined as follows:
$$
J(A, B)=\frac{|A \cap B|}{|A \cup B|}
$$
Then \(Sim_{io}(F_1, F_2)\) can be calculated as:
$$
Sim_{io}(F_1,F_2) = J(IO_{F_1}, IO_{F_2})= \frac{|IO_{F_1} \cap IO_{F_2}|}{|IO_{F_1} \cup IO_{F_2}|}
$$

\noindent
{\bf Callees}, which represent the dependencies of functions, also serve as indicators of functional similarity between code blocks~\cite{mcmillan2012detecting,novak2019source}.
Similar to input and output arguments, the static analyzer collects and formalizes callees of \(F_1\) as a set \(Call_{F_1}=\{f^1_1, f^1_2, ....,f^1_n\}\), where \(f^1_i\) denotes a callee of \(F_1\).
\sys also computes \(Sim_{call}(F_1, F_2)\) through Jaccard similarity coefficient.
$$
Sim_{call}(F_1,F_2) = \frac{|Call_{F_1} \cap Call_{F_2}|}{|Call_{F_1} \cup Call_{F_2}|}
$$



\begin{figure}
    \centering     
    \includegraphics[width=0.85\textwidth]{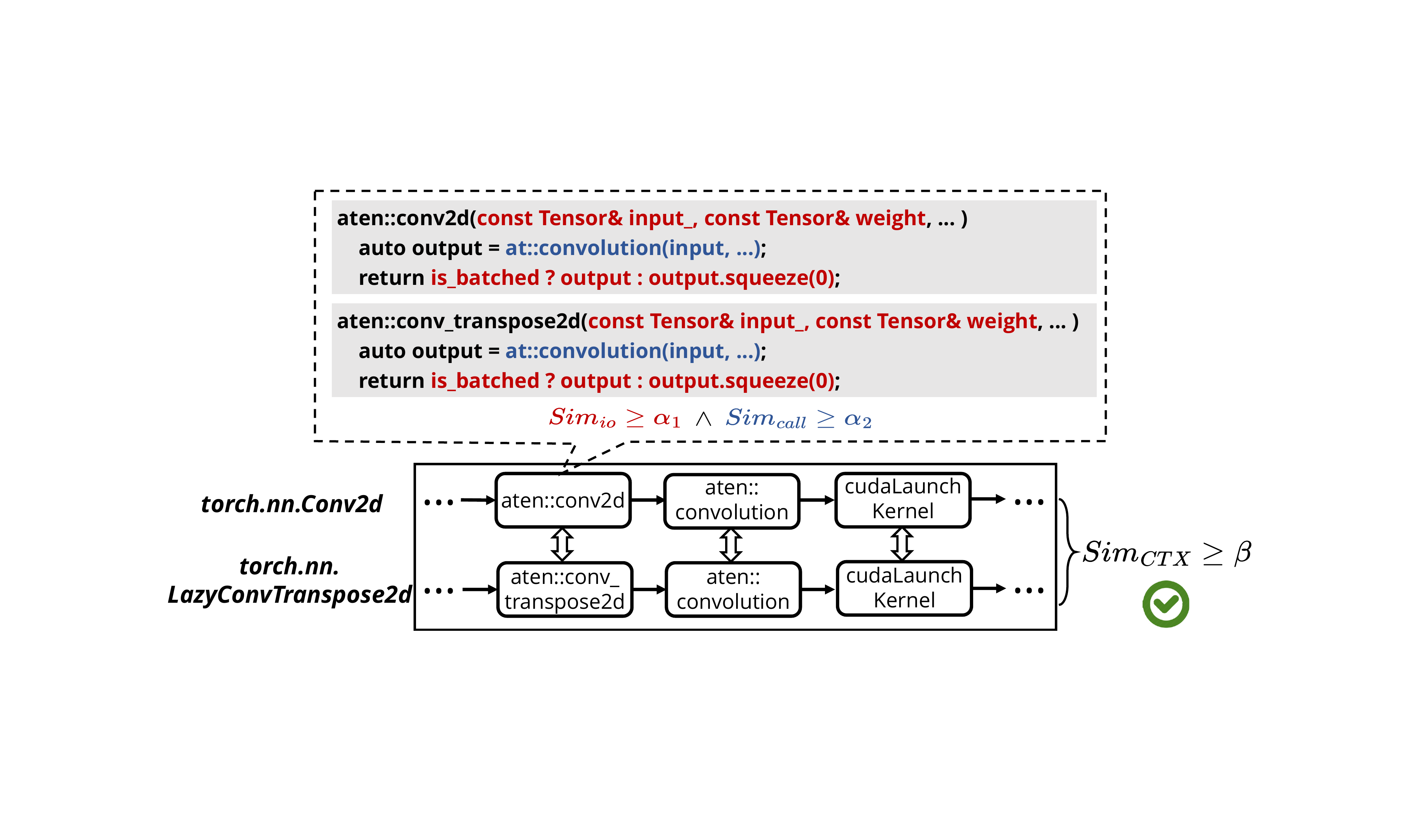}
    \caption{A Demo Case of Matching APIs Pairs with Context Information}
    \label{fig:dynamic}
\end{figure}


The static analyzer considers two source code functions to have similar implementations when both similarities exceed the built-in threshold.
$$
Sim_{io}(F_1,F_2) \geq \alpha_1 \land Sim_{call}(F_1,F_2) \geq \alpha_2,
$$
otherwise, they are considered dissimilar.
The dashed box in~\autoref{fig:dynamic} shows how the static analyzer calculates the similarity between two source code functions \texttt{aten::conv2d} and \texttt{aten::conv\_transpose2d}.
Red marks input and output arguments, and blue highlights callees.
Given that these two functions share identical input and output arguments (\eg,  \texttt{const Tensor\& weight}) and call the same functions (\eg,  \texttt{at::convolution}), both \(Sim_{io}\) and \(Sim_{call}\) exceed the thresholds.
Therefore, \texttt{aten::conv2d} and \texttt{aten::conv\_transpose2d} are classified into one group of analogous source code functions.

\subsection{API matcher}\label{sec:apimatcher}
\begin{figure}
    \centering     
    \includegraphics[width=0.7\textwidth]{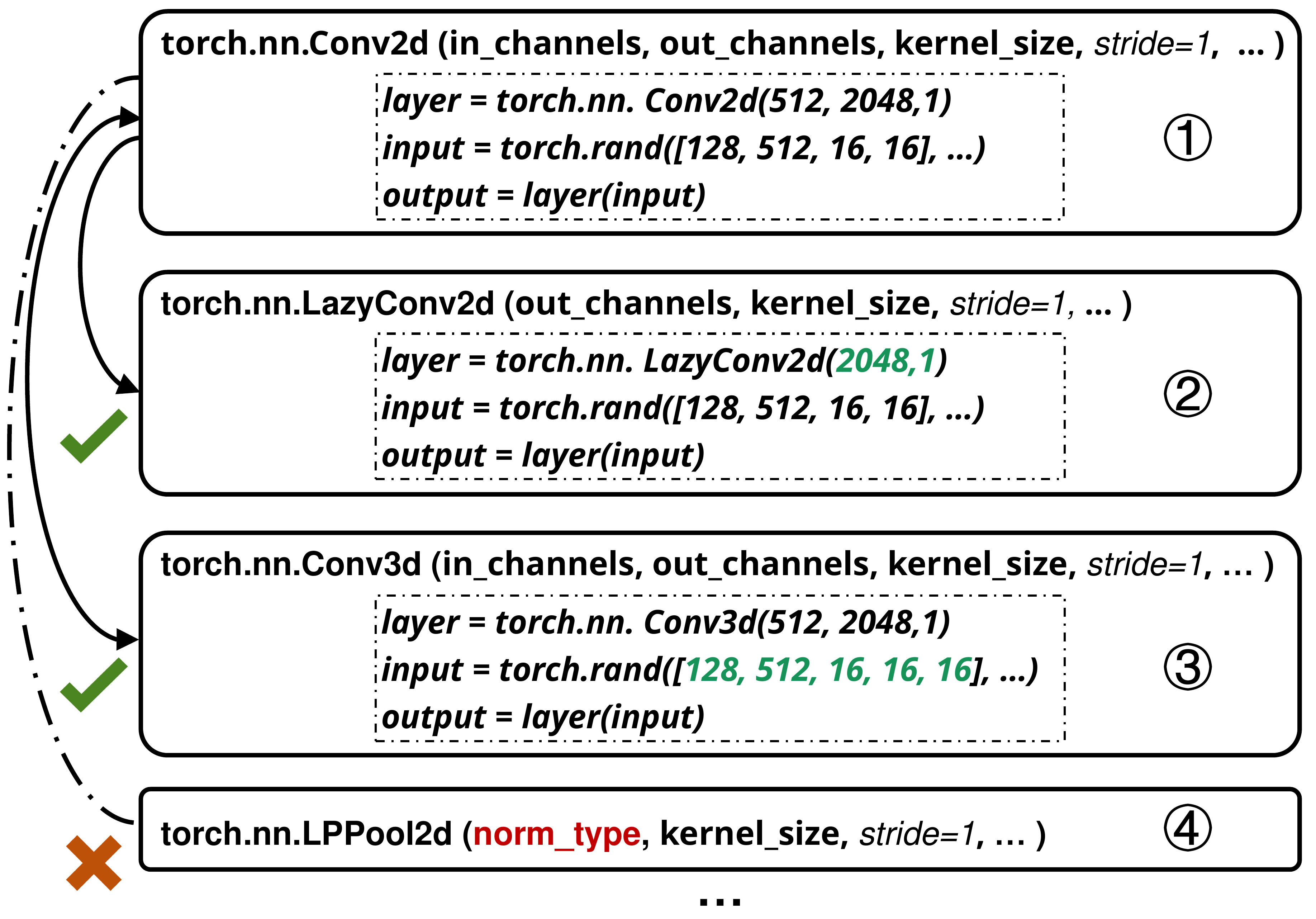}
    \caption{Verifying API Pairs and Generating Cases }\label{fig:verify}
\end{figure}

The API matcher first receives context information from the dynamic profiler and static analyzer (\ie, analogous function groups and API call stacks) and matches analogous API pairs based on context similarity and signature similarity~\cite{deng2022fuzzing}.
Subsequently, it checks the arguments of analogous API functions and discards the API pairs with unsolvable argument mismatches, which renders the buggy code unusable in test case generation.
For instance, the API functions \texttt{Conv2d} and \texttt{LPPool2d} each have required arguments that are not included in the other, making \sys unable to generate a test case for one based on the code of the other, as shown in~\autoref{fig:verify} (\WC{4}).
Consequently, this API pair is considered to have encountered an argument mismatch and is discarded.
Note that matching and filtering API pairs in API matcher is a one-time process.
The matched analogous API pairs can be saved and reused in the test case generator.

\noindent
\textbf{Matching.}
For an API function \(A_S\), the call stacks obtained from the dynamic profiler include a set of source code functions it calls during execution.
The static analyzer indicates that some of these functions share similar functionality and implementations to other source code functions.
The API matcher integrates these two parts of context information to obtain the execution context \(CTX_S\) of the API function:

$$
CTX_S = \{f'_1, f'_2, ..., f'_m\} \bigcup \{f^S_1, f^S_2, ..., f^S_n\}
$$
where \(f'_i\) indicates the source code functions in the groups identified in the static analyzer and \(f^S_j\) represents other source code functions in the call stack that do not belong to any of the groups.
Similarly, we denote the execution context of another API function \(A_T\) as \(CTX_T\).
\upd{Response to R2Q1: }{
Note that \sys does not merely compare whether the two API call stacks are the same.
When calculating the \(Sim_{CTX}(A_S,A_T)\), if both APIs call source code functions from the same group, these functions will be treated as one function because of their similar functionality and implementations.
Such a design enables \sys to match APIs with similar underlying implementations but different call stacks.
}
The context similarity between \(A_S\) and \(A_T\) can be calculated by the Jaccard similarity coefficient:
$$
Sim_{CTX}(A_S,A_T) =\frac{|CTX_S \cap CTX_T|}{|CTX_S \cup CTX_T|}
$$

\sys uses the threshold \(\beta\) to evaluate the context similarity between \(A_S\) and \(A_T\) and considers the two APIs are similar when \(Sim_{CTX}(A_S, A_T)\) exceeds \(\beta\).
\autoref{s:configurable} explains the selection of the default value of \(\beta\) in detail.


\upd{Response to R2Q1: }{
\autoref{fig:dynamic} illustrates a simplified process of matching \texttt{Conv2d} and \texttt{LazyConvTranspose2d} mentioned in our motivation example via context information.
Based on the records of the dynamic profiler (depicted in the solid box), both APIs call the source code functions \texttt{aten::convolution} and \texttt{cudaLaunchKernel}.
For source code functions \texttt{aten::conv2d} and \texttt{aten::conv\_transpose2d}, which are separately called by \texttt{Conv2d} and \texttt{LazyConvTranspose2d}, they have been divided into one analogous function group by the static analyzer and are considered as the same function when calculating \(Sim_{CTX}\).
Note that there are multiple differences in the call stacks of \texttt{Conv2d} and \texttt{LazyConvTranspose2d}, \texttt{aten::conv2d} and \texttt{aten::conv\_transpose2d} are two demo cases.
}
Since both APIs call almost the same source code functions, their Jaccard similarity coefficients are greater than \(\beta\), and we match them as a context-similar API pair.
In addition to context-similar API pairs, we supplement analogous API pairs matched by the signature similarity calculated by DeepREL~\cite{deng2022fuzzing}, which are publicly available.
Based on their experiment results, we select the top 20 API functions with the highest signature similarity to one API as its analogous API functions.

\noindent
\textbf{Filtering.}
API matcher checks arguments of matched API functions to avoid argument mismatch problems in the test case generator.
For the source API \(A_S\), its arguments \(P_S\) can be represented as:
\(P_S = P^r_S \bigcup P^o_S\),
where \(P^r_S\) refers to required arguments and \(P^o_S\) refers to optional arguments.
\sys discards the API pair \((A_S, A_T)\), iff \(A_S\) and \(A_T\) each contain required arguments \(p^r_i\) and \(p^r_j\) that are not included in the other's arguments set, which means that the test case of either API cannot provide values of required arguments to the other API and generate new test cases.

The discarded API pairs satisfy the following:
$$
(\exists p^r_i \in P^r_S, p^r_i \notin P_T) \wedge (\exists p^r_j \in P^r_T, p^r_j \notin P_S)
$$



\autoref{fig:verify} shows an example of verifying API arguments, where \(A_S\) is \texttt{torch.nn.Conv2d}(\WC{1}).
Its required argument set \(P^r_S\) includes `in\_channels', `out\_channels',  and `kernel\_size', and the optional argument set \(P^o_S\) contains `stride' (default value is 1), etc.
One of its analogous APIs, \texttt{torch.nn.Conv3d}, has the same argument set, allowing the two APIs to generate test cases for each other in the generator and pass the verification(\WC{3}).
\texttt{torch.nn.Lazyconv2d} has required arguments `out\_channels' and `kernel\_size'.
The test code of \texttt{Conv2d} can be modified by removing the first argument to generate test cases for the target API \texttt{LazyConv2d}, which also passes the verification(\WC{2}).
Unfortunately, \texttt{torch.nn.LPPool2d} has a required argument `norm\_type' that are not present in \texttt{Conv2d}, and \texttt{LPPool2d} lacks required arguments `in\_channels' and `out\_channels' (\WC{4}).
Due to the lack of values of required arguments (\ie, `in\_channels', `out\_channels', and `norm\_type'), \sys cannot generate new test cases for either API based on the test cases of the other, therefore \sys discards the API pair that consists of \texttt{LPPool2d} and \texttt{Conv2d}.

\subsection{Test Case Generator}\label{sec:generate}

Given a verified API pair of \(A_S\) and \(A_T\), the test case generator synthesizes new test cases \(C_T\) for the target API based on the collected buggy code \(C_S\) of the problematic API \(A_S\) (\ie, source API).
As shown in the case of~\autoref{fig:moti}, the generator can adaptively adjust the code of \(C_T\) to resolve two kinds of differences (if any) between APIs, namely argument difference and dimension difference, thereby ensuring that the newly generated test cases are executable.

\noindent
\textbf{Argument Difference.}
When the argument set of the source API \(A_S\) includes arguments not present in the target API \(A_T\) (\eg, \texttt{Conv2d} and \texttt{LazyConv2d} in~\autoref{fig:verify}), an argument difference arises.
To solve this problem, the test case generator modifies the test case \(C_T\) by removing irrelevant arguments to make it executable for \(A_T\).
The dashed box in \autoref{fig:verify} provides an example of resolving arguments difference (\WC{2}).
The argument set of the source API \texttt{Conv2d} contains the first argument `in\_channels' that the target API \texttt{LazyConv2d} does not have.
Therefore, the generator discards the value `512' corresponding to the first argument and keeps only the values `2048' and `1' corresponding to other arguments (marked by green).
Additionally, \autoref{fig:moti} provides another example that removes the argument `in\_channel=512' in generating test cases for \texttt{LazyConvTranspose2d}.

\noindent
\textbf{Dimension Difference.}
As mentioned previously, the DL framework provides a series of APIs for inputs of varying dimensions (e.g, \texttt{Conv2d} and \texttt{Conv3d}), typically sharing similar implementations and susceptibility to similar bugs~\cite{torch_issue1}.
However, the existing methods encounter challenges in constructing test cases for these API functions due to the different dimensions of their arguments~\cite{deng2022fuzzing}.
To resolve the dimension difference, the test case generator first obtains the API signatures through open-source libraries (\eg, the `inspect' library in Python) and identifies the dimension-related arguments from signatures.
It then dynamically adjusts the test code by increasing or decreasing the dimensions of argument values based on the dimension information of the API signatures of the source and target APIs, and generates test cases for those API functions.
\autoref{fig:verify} shows an example of resolving dimension difference and generating available test cases for \texttt{Conv3d} (\WC{3}).
The API signature shows that the `input' of \texttt{Conv3d} and \texttt{Conv2d} is dimension-related, and the `input' of \texttt{Conv2d} is a 4-dimensional tuple, and the `input' of \texttt{Conv3d} should be a 5-dimensional tuple.
The generator recognizes such a dimension difference and expands the 4-dimension tuple received by \texttt{Conv2d} to a 5-dimension tuple to adjust the input dimensions (marked in green) and generates executable test cases for \texttt{Conv3d}.

\sys adaptively generates test cases for various target APIs based on the bug detected in the source API \texttt{Conv2d} and finally identifies status bugs on \texttt{Conv3d} and \texttt{LazyConv2d}, which has been reported to the developers~\cite{crashbug1}.

\subsection{Test Case Evaluator}\label{sec:evaluator}
Existing work usually considers the execution results of another API function as a pseudo test oracle and checks whether two API functions produce equal results to detect potential bugs~\cite{deng2022fuzzing,guo2020audee,pham2019cradle}.
However, they cannot detect performance bugs due to the difficulty of obtaining test oracles for the runtime overhead of APIs.
To effectively identify API bugs regardless of bug types, the evaluator considers the buggy behavior of the source test case \(C_S\) as the test oracle and observes whether the new test case \(C_T\) has the same buggy behavior.
Specifically, the evaluator identifies three types of bugs on \(A_T\) as follows.

\noindent
\(\bullet\)
\textit{Status bug} arises when \(C_T\) throws the identical exception as the source case \(C_S\).
The evaluator collects exception details, and if \(C_T\) throws the same exception as the original bug case \(C_S\), it is considered that \(A_T\) has a status bug.

\noindent
\(\bullet\)
\textit{Value bug} arises when \(C_T\) generates the same specific numerical errors (\eg, NaN) as \(C_S\).
The evaluator logs the outputs of test cases,
and if \(C_T\) produces anomalous values matching those described in the bug report of \(A_S\), \(A_T\) is deemed to have a value bug.

\noindent
\(\bullet\)
\textit{Performance bug} arises from an underlying implementation or optimization error, leading to an unexpectedly high overhead on APIs.
To detect performance bugs, \sys calculates the expected overhead by leveraging the bug report of \(A_S\) and records the actual runtime overhead.
If \(C_T\) exhibits unexpected overhead identical to that described in the bug report of the problematic API \(A_S\) (\eg, a time overhead greater than expected), \(A_T\) is considered to have a performance bug.
Take the case in~\autoref{fig:moti} as an example, the bug report of \texttt{Conv2d} indicates that the performance bug causes the group convolution to exhibit a greater time cost than implementing these convolutional layers individually, which is confirmed by the developers.
Leveraging such an anomalous behavior as the test oracle,
\sys reveals that \texttt{LazyConvTranspose2d} also experiences a higher time overhead under the group setting than implementing \texttt{LazyConvTranspose2d} layers individually, thereby identifying the performance bug.
Please note that this does not imply that \texttt{LazyConvTranspose2d} incurs the same cost as \texttt{Conv2d} in the original bug report, but rather that both achieve costs exceeding the cost of individual execution (i.e., the expected overhead).

\section{Evaluation}\label{sec:eval}

In this section, we aim to answer the following research questions.

\noindent \textbf{RQ1:} How effective is \sys in detecting real-world bugs?

\noindent \textbf{RQ2:} How efficient is \sys in detecting real-world bugs?

\noindent \textbf{RQ3:} What is the impact of different configurable parameters in \sys?

\subsection{Setup}
\label{s:setup}

\noindent
{\bf Baseline and Metric:}
We use three state-of-the-art open-sourced testing tools for comparison, namely DocTer~\cite{xie2022docter}, DeepREL~\cite{deng2022fuzzing} and TitanFuzz~\cite{deng2023large}.
For the data not shown in their paper (\eg, the number of cases generated by a complete execution), we obtain it by running their open-source code.
\sys mainly compared with baselines from three metrics:

\noindent
\(\bullet\)
\textit{Ratio of test cases that can trigger bugs.}
This metric is calculated by dividing the number of cases that can trigger or expose bugs by the total number of generated cases, which can reflect the efficiency of a test approach in generating test cases and detecting bugs.
Note that if the baselines do not provide the number of valid cases that can trigger bugs, we will use the number of all bug candidate cases (\ie, the upper bound of valid cases, which may contain a significant number of duplicates) to estimate this metric for the baselines, although this may overestimate the baselines' results on this metric.

\noindent
\(\bullet\)
\textit{Average time to detect bugs.}
Following the prior work~\cite{cheu2003support,zhao2021identifying}, we use the metric of average time to detect bugs to compare the bug detection efficiency of each method.
\upd{Response to R1Q1: }{
Specifically, we record the time taken by each method to conduct a complete test run and subsequently divide it by the number of detected bugs in the execution.
Note that this metric explicitly excludes any preprocessing time, such as API matching in CITADEL and DeepREL or constraint construction in DocTer, since these steps often involve manual effort or produce static results that can be reused in multiple tests.
}
Since the manual cost of verifying which thousands of cases generated by baselines are real bugs is too high, for the baseline method, we directly use the total number of bugs reported in their papers.
Note that the fuzzy-based baselines can explore more behaviors and trigger bugs through multiple executions.
We actually calculate the upper bound of baselines' performance in this metric, which is the time of a complete execution divided by the total number of bugs they report.
In contrast, \sys exploits existing bug reports to discover API bugs without fuzzy generation, and all bugs reported by \sys are discovered in one execution.

\noindent
\(\bullet\)
\textit{Number of covered APIs.}
In prior work, DocTer~\cite{xie2022docter} reports the number of APIs that successfully extract constraints as the number of covered APIs, and DeepREL~\cite{deng2022fuzzing} counts APIs invoked by their `API\_Match\_Verifier', which are successfully included in equivalent pairs.
Following the prior work, we report the number of APIs matched and verified in our API matcher.
In addition, we also count the number of analogous APIs covered by the collected bug cases used in our experiments.

\noindent
{\bf Collected Issues and Context Information.}
In the experiment, we separately collect and verify 258 and 288 valid bug cases from the latest 30,000 issues of PyTorch and the latest 20,000 issues of TensorFlow, as one input to \sys.
This process takes approximately 3 weeks per participant.
Among them, the problematic APIs in 172 cases (\ie, 104 PyTorch cases and 68 TensorFlow cases) have matched analogous APIs and are used to generate new test cases in the test case generator.
These bug cases are publicly available in our repository~\cite{our_repo}.
In addition, the dynamic profiler constructs test cases for 999 PyTorch APIs and 2,076 TensorFlow APIs and records their call stacks.
Based on the experimental results of prior work~\cite{su2016identifying,misu2017interface}, we set \(\alpha_1=0.8\) as the threshold of input and output arguments similarity.
To strictly judge analogous source code functions and reduce the impact of false positives on subsequent API matching, we set \(\alpha_2=0.8\) as the threshold for callees similarity. 
If both the arguments similarity and callees similarity of two functions exceed the thresholds, the two source code functions are judged as analogous functions.
The static analyzer separately selects 944 and 7,028 functions in PyTorch and TensorFlow source code that share similarity with at least one other function from the source code and divides them into 2,467 groups.
The call stacks and analogous function groups are input into the API matcher as context information.

\noindent
{\bf Software and Hardware:}
The prototype of \sys is implemented on top of Python 3.9. 
In our experiments, \sys test and identify bugs across PyTorch 1.7.0 to 1.13.1 and TensorFlow 2.1.0 to 2.13.0.
All experiments are conducted on a server with Intel(R) Xeon(R) Gold 6226R 2.90GHz 16-core processors, 130 GB of RAM, and an NVIDIA 3090 GPU running on Ubuntu 22.04.

\subsection{Effectiveness in Detecting Bugs}\label{s:effective}



\begin{table}[]
    \caption{Summary of Detected API Bugs on PyTorch and TensorFlow}\label{tab:rq1-1}
    \centering
    \footnotesize
    \tabcolsep=1pt
    \begin{tabular}{ccccccccccccccccccccc}
    \toprule
    \multirow{2}{*}{Framework} & \multicolumn{4}{c}{\#Total} & \multicolumn{4}{c}{\#Rejected} & \multicolumn{4}{c}{\#Duplicated} & \multicolumn{4}{c}{\#New} & \multicolumn{4}{c}{\#Confirmed} \\ \cmidrule(r){2-5}\cmidrule(r){6-9}\cmidrule(r){10-13}\cmidrule(r){14-17}\cmidrule(r){18-21}
     & Total & Stat . & Val . & Perf . & Total & Stat . & Val . & Perf . & Total & Stat . & Val . & Perf . & Total & Stat . & Val . & Perf . & Total & Stat . & Val . & Perf . \\ \midrule
    PyTorch & 77 & 52 & 15 & 10 & 1 & 0 & 1 & 0 & 18 & 16 & 2 & 0 & 58 & 36 & 12 & 10 & 36 & 21 & 7 & 8 \\
    TensorFlow & 74 & 51 & 20 & 3 & 2 & 2 & 0 & 0 & 6 & 6 & 0 & 0 & 66 & 43 & 20 & 3 & 56 & 36 & 17 & 3 \\ \midrule
    Total & 151 & 103 & 35 & 13 & 3 & 2 & 1 & 0 & 24 & 22 & 2 & 0 & 124 & 79 & 32 & 13 & 92 & 57 & 24 & 11 \\ \bottomrule
        \end{tabular}
\end{table}


\noindent
{\bf Experiment Design:} 
To evaluate the effectiveness of \sys in detecting real-world bugs, we conduct experiments on the PyTorch and TensorFlow frameworks and report all detected API bugs to developers for confirmation.
Our experiment counts the number of API bugs detected by \sys, that is, when calling an API with certain inputs triggers one bug, \sys will consider that an API bug is detected.
In addition, aligned to prior work~\cite{xie2022docter}, 
different inputs triggering the same unexpected behavior on one API will only be counted once.
During this process, we record the states of reports, such as confirmed or rejected, and the bug types.
Furthermore, following the setting of previous work~\cite{deng2022fuzzing,wei2022free}, we collect the number of verified pairs and covered APIs in the API matcher of \sys.
Note that \sys uses the test code in existing bug reports to generate new test cases, and the number of APIs tested in the experiment is related to the collected bug cases.
We also report the number of tested APIs using the collected 172 bug cases.
How to collect more bug cases to fully utilize the matched API pairs and test more APIs will be a future direction.

\begin{figure}
    \centering     
    \includegraphics[width=0.7\textwidth]{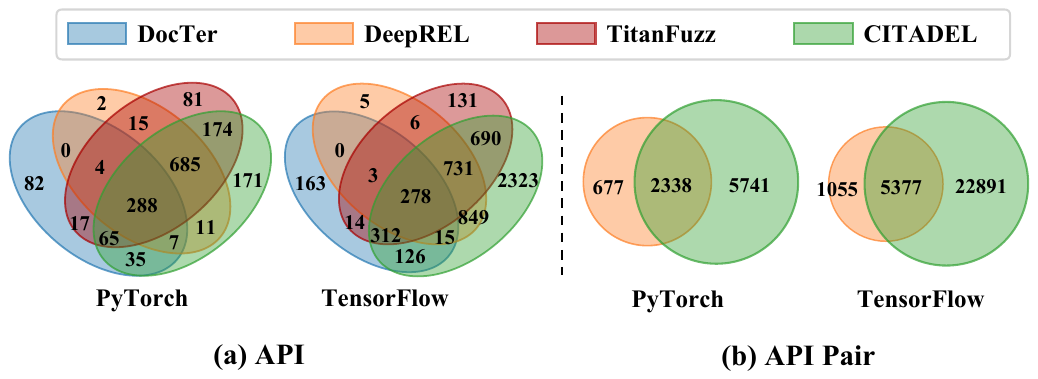}
    \caption{Comparison of API Coverage and Matched API Pairs}
    \label{fig:rq2}
\end{figure}

\noindent
{\bf Results:}
\autoref{tab:rq1-1} summarizes the three types of bugs detected by \sys, namely `Stat.', `Val.', and `Perf.' (\ie, status, value, and performance bugs).
\updmj{Response to R2Q2: }{
Following the prior work~\cite{xie2022docter,deng2022fuzzing,deng2023large}, the number of bugs reported here is the number of abnormal behaviors in different APIs (i.e., API bugs), rather than the number of independent implementation defects.}
The first column of~\autoref{tab:rq1-1} displays the DL framework and the following columns indicate the number of all detected API bugs (`\#Total'), API bugs that developers do not plan to work on (`\#Rejected'), API bugs that have been reported in existing reports (`\#Duplicated'), new API bugs that have not been reported (`\#New'), new API bugs that have been confirmed (`\#Confirmed').
In addition,~\autoref{tab:rq1-2} presents a comparison between \sys and baselines in the three metrics in~\autoref{s:setup}.
The first column shows four test approaches in comparison and the second column displays the DL framework.
The columns `\#API' and `\#Pairs' show the number of covered APIs and matched API pairs in each approach, corresponding to the `\textit{Number of covered APIs}' in~\autoref{s:setup}.
In addition, \autoref{fig:rq2} uses Venn diagrams to present the comparison of the number of covered APIs and matched API pairs between \sys and the baselines.
Since we could not find specifics on the APIs and pairs covered by DeepREL in its repository, we use the data collected from its full execution.


\begin{table}
    \caption{\upd{Response to R3Q2: }{Comparison of \sys and Baselines Result \footnotesize{(Brackets mark the APIs and API Pairs Covered by the Collected Bug Cases)}}}
    \label{tab:rq1-2}
    \centering
    \footnotesize
    \tabcolsep=3pt
    \begin{tabular}{cccccccc}
    \toprule
    \multirow{2}{*}{Approach} & \multirow{2}{*}{Framework} & \multicolumn{2}{c}{API Coverage} & \multicolumn{3}{c}{Case Generation} & \multirow{2}{*}{\begin{tabular}[c]{@{}c@{}}Average Time To\\ Detect Bugs (min)\end{tabular}} \\ \cmidrule(r){3-4} \cmidrule(r){5-7}
    &  & \#API & \#Pairs & \#Valid & \#Total & Ratio (\%) &  \\ \midrule
    \multirow{3}{*}{DocTer} & PyTorch & 498 & \textbackslash{} & 45 & 17,227 & 0.26 & 107.16 \\
        & TensorFlow & 911 & \textbackslash{} & 206 & 16,632 & 1.24 & 25.98 \\ \cmidrule{2-8}
        & Total & 1,409 & \textbackslash{} & 251 & 33,859 & 0.74 & 41.98\\ \midrule
    \multirow{3}{*}{DeepREL} & PyTorch & 1,071 & 4,290 & 2,001 & 77,662 & 2.58 & 40.63\\
        & TensorFlow & 1,902 & 8,808 & 2,052 & 252,533 & 0.81 & 64.71 \\ \cmidrule{2-8}
        & Total & 2,973 & 13,098 & 4,053 & 330,195 & 1.23 & 58.62 \\ \midrule
    \multirow{3}{*}{TitanFuzz} & PyTorch & 1,329 & \textbackslash{} & 2,406	 & 158,185 & 1.52 & 43.05\\
        & TensorFlow & 2,215 & \textbackslash{} & 11,235 & 191,862 & 5.86 & 101.96 \\ \cmidrule{2-8}
        & Total & 3,544 & \textbackslash{} & 13,641 & 350,047 & 3.90 & 68.43\\ \midrule
    \multirow{3}{*}{\sys} & PyTorch & 1,436 (529) & 8,079(797) & 82 & 196 & 41.84 & 4.83 \\
        & TensorFlow & 5,380 (675) & 28,268 (1,387) & 61 & 208 & 29.33 & 6.61 \\ \cmidrule{2-8}
        & Total & 6,816 (1,204) & 36,347 (2,184) & 143 & 404 & 35.40 & 5.70 \\ \bottomrule
    \end{tabular}
\end{table}

\noindent
{\bf Analysis:}
The results in~\autoref{tab:rq1-1} illustrate the effectiveness of \sys in detecting various types of real-world bugs.
\sys generates test cases based on a total of 172 real bugs collected from GitHub repositories and successfully detects 151 API bugs on PyTorch and TensorFlow, out of which only 3 are rejected by developers.
These cases show the same anomalous behaviors (\eg, NaN) as the source problematic API, but developers have no plans to fix them.
The following `Rejected Case' provides a detailed analysis of a rejected case.
Of the remaining 148 bugs, 24 are duplicates of existing bug reports and the remaining 124 are unreported API bugs, 92 of which have been confirmed by developers.
Excluding rejected cases, \sys detected 101 status bugs, 34 value bugs, and 13 performance bugs, demonstrating its effectiveness in detecting different types of DL framework bugs.
\updmj{Response to R2Q2: }{
Furthermore, we analyze the patches provided by developers in response to our bug reports and count the number of API bugs that share the same patch as the source API bug.
In PyTorch, eight of our issue reports receive official patches (including twelve API bugs), while the patches in six reports indicate that seven newly reported API bugs share the same underlying implementation errors with their corresponding source bugs.
In addition, due to the inactivity of the TensorFlow community, none of our bug reports have received patches, making this analysis impossible.
This finding highlights a characteristic of our context similarity-based method in \sys, i.e., it could identify multiple API bugs caused by the same underlying implementation error.
}
Moreover, compared to other types of bugs, the number of detected performance bugs is relatively small.
Our manual analysis of the PyTorch and TensorFlow repositories reveals that the reported performance bugs are infrequent.
Take the PyTorch framework as an example,
only approximately 300 issues are labeled as `performance' out of over 10,000 open issues.
Moreover, the limited number of issues with reproducible code poses a challenge for \sys to gather a significant amount of code related to performance bugs in DL framework repositories.
Finally, 8/172 real bugs that \sys collects to generate test cases for the matched API pairs are related to performance.
Based on these collected performance bugs, \sys detects 13 new performance bugs, and 11 of them have been confirmed.


The API matcher of \sys covers a total of 1,436 PyTorch APIs and 5,380 TensorFlow APIs, which is 365 more PyTorch APIs and 3,478 more TensorFlow APIs than DeepREL's `API\_Match\_Verifier' and significantly more than 498 PyTorch APIs and 911 TensorFlow APIs covered by DocTer's constraints.
\autoref{fig:rq2} show the comparison of API and API pairs covered by \sys and baselines, respectively.
\upd{Response to R3Q2: }{
In addition, the experiments based on the collected bug cases cover 529 PyTorch APIs and 675 TensorFlow APIs, among which 221 and 338, respectively, are not covered by the baseline methods.
\sys successfully identifies one PyTorch bug and eight TensorFlow bugs on these previously uncovered APIs.
Furthermore, the corresponding test cases cover 797 and 1,387 analogous API pairs on PyTorch and TensorFlow, of which 574 and 1,154 are not covered by the baselines.
Note that, due to the limited number of collected bug cases, the experiment currently only uses a small subset of matched analogous APIs and API pairs. Exploring techniques to automatically annotate and efficiently collect a broader set of bug cases to further enhance the detection effect of \sys will be our future research direction (\autoref{sec:discuss}).
}
The newly covered API pairs enable \sys to successfully detect 49 API bugs on PyTorch and TensorFlow, comprising 41 status bugs, 3 value bugs, and 5 performance bugs,
which demonstrates the effectiveness of the newly covered APIs and API pairs of \sys in detecting bugs.
For example, \sys newly matches the API pair of \texttt{Conv2d} and \texttt{LazyConvTranspose2d} via the context similarity and identifies the performance bug in \texttt{LazyConvTranspose2d} (\autoref{fig:moti}).
The API pair of \texttt{ReLU6} and \texttt{HardTanh} in the following case study is also newly covered in \sys.



\begin{figure}
    \centering     
    \includegraphics[width=0.65\textwidth]{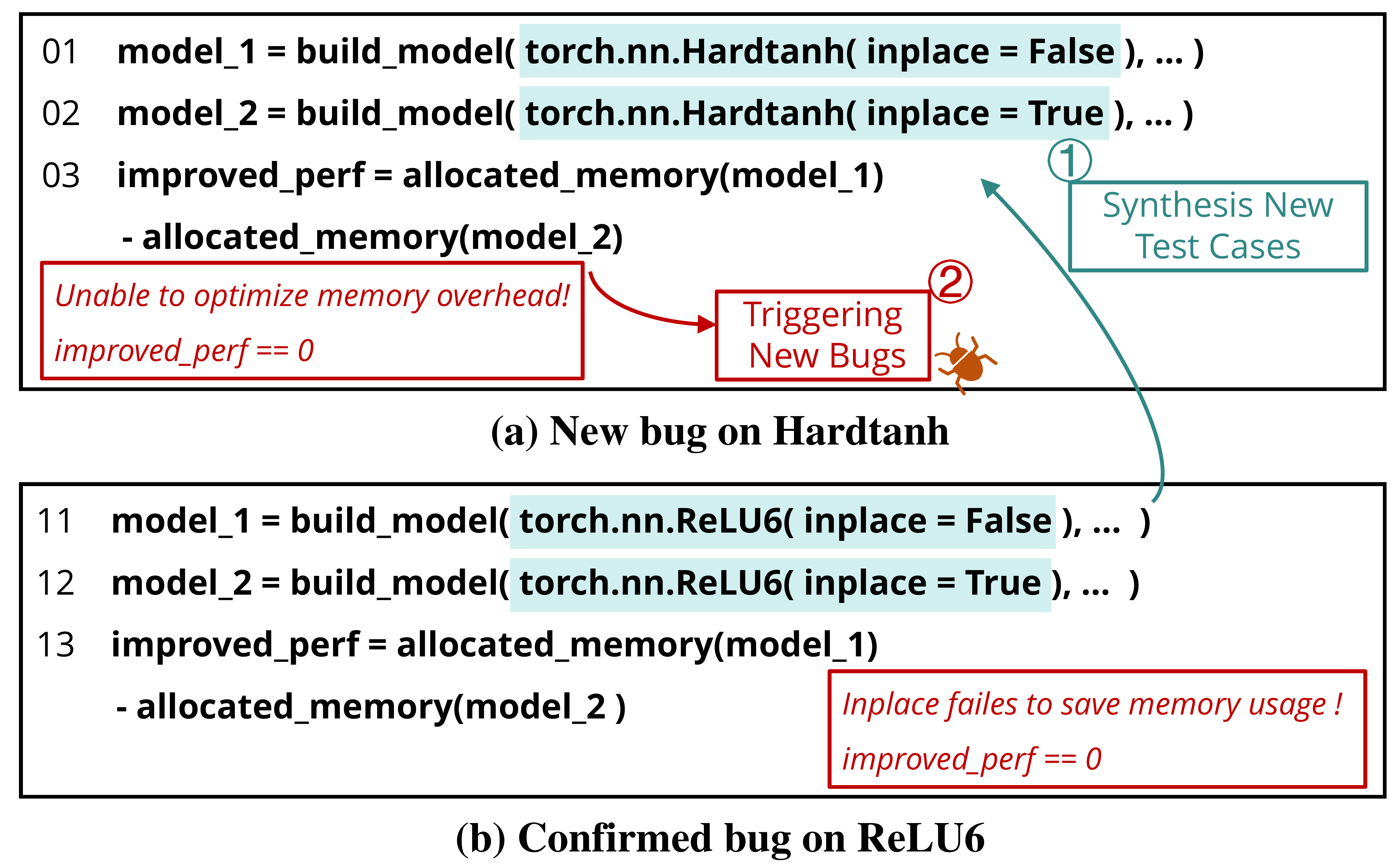}
    \caption{Performance Bug on \texttt{Hardtanh}}\label{fig:rq1-case1}
\end{figure}

\noindent
\textbf{Bug Case 1:}
In the PyTorch 1.13.1 release, \sys detects a performance bug on \texttt{Hardtanh} API function, and~\autoref{fig:rq1-case1} presents how \sys identify this performance bug.
When this bug occurs, the `inplace' argument in \texttt{Hardtanh} can not optimize the memory overhead, and no matter `inplace' is assigned as `True' or `False', the GPU memory allocated by the model with \texttt{Hardtanh} remains constant, as shown in Lines 1-3 in~\autoref{fig:rq1-case1}(a).
Nevertheless, in the 1.8.0 and 1.9.0 versions, this argument can effectively decrease memory usage (\eg, reduce from 40.43 MB to 21.82 MB).
To discover this performance bug, \sys first matches \texttt{ReLU6} and \texttt{Hardtanh} as analogous API functions.
Then, based on a collected bug on \texttt{ReLU6}, \sys synthesizes and executes a test case for \texttt{Hardtanh} (\WC{1}).
The bug report of \texttt{ReLU6} shows that the model with \texttt{ReLU6} allocates the same amount of GPU memory regardless of whether `inplace' is enabled, and the variable `improved\_perf' in Line 13 of~\autoref{fig:rq1-case1}(b) is zero.
\sys leverages the anomalous behavior of enabling the `inplace' not decreasing the memory overhead as the test oracle and generates a new test case, as shown in Line 3 of~\autoref{fig:rq1-case1}(a). 
Finally, \sys identifies the performance bug on \texttt{Hardtanh} (\WC{2}), and the developers have confirmed this bug and labeled it as `high priority'~\cite {perfbug2}.

\begin{figure}
    \centering     
    \includegraphics[width=0.75\textwidth]{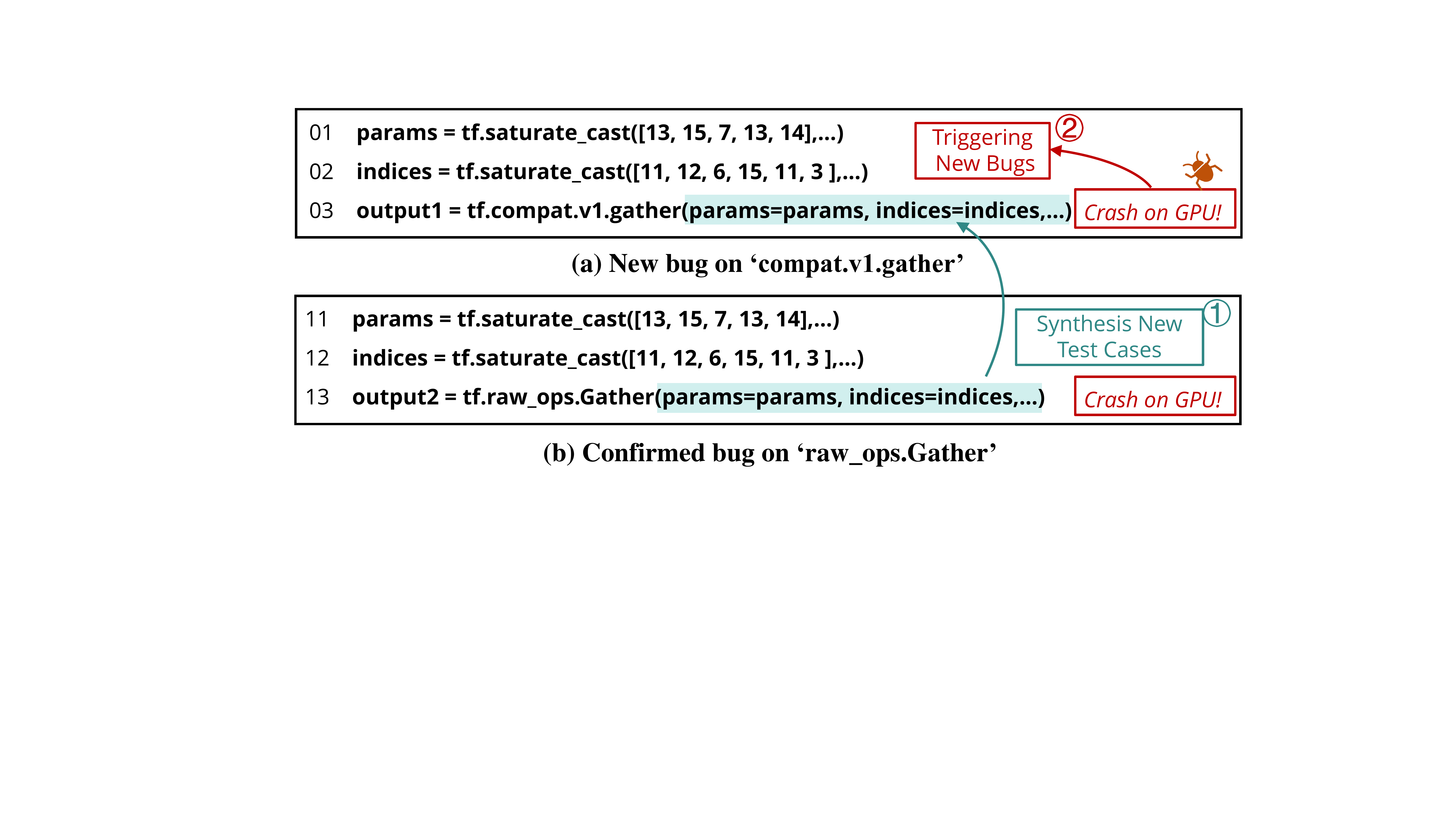}
    \caption{Crash on \texttt{tf.compat.v1.gather}}\label{fig:rq1-case2}
\end{figure}

\noindent
\textbf{Bug Case 2:}
\sys detects a status bug on the \texttt{tensorflow.compat.v1.gather} in TensorFlow 2.14.0 release.
When the last dimension of the `params' argument takes a specific value (\eg, 14), \texttt{tf.compat.v1.gather} will crash directly on the GPU without throwing any error message.
To detect this status bug, \sys first matches \texttt{tensorflow.compat.v1.gather} with the source API \texttt{tensorflow.raw\_ops.Gather} based on context similarity (over 0.9).
Subsequently, \sys analyzes the arguments of the two API functions and synthesizes a new test case using the collected bug case on the source API for the analogous API \texttt{v1.gather}, as shown in~\autoref{fig:rq1-case2} (\WC{1}).
The new test case on \texttt{v1.gather} has crashed, exhibiting the same anomalous behavior as observed in \texttt{raw\_ops.Gather} (\WC{2}).
Finally, \sys identifies the new status bug on \texttt{v1.gather}.
The newly discovered bug has been reported and confirmed by developers~\cite{crashbug2}.

\noindent
\textbf{Rejected Case:}
Although \sys effectively identifies API bugs based on whether analogous APIs exhibit the same anomalous behavior as the problematic API, several cases are still rejected by developers.
Note that, these rejected cases still exhibit anomalous behaviors, but developers consider them unimportant and have no plans to fix or work on them.
Here, we present an example.
\sys encounters a rejected case when detecting a bug on the \texttt{torch.logdet} API. 
During testing, \sys matches \texttt{torch.logdet} with \texttt{torch.det} and then constructs test cases for the target API \texttt{logdet} based on a buggy code of the source API \texttt{det}.
When executing the test cases, \sys finds that both APIs have abnormal and dangerous output value `NaN' (\ie, not a number) which could further affect subsequent calculations and raise dangerous behaviors~\cite{odena2019tensorfuzz}.
However, developers suggest that the abnormal `NaN' output on \texttt{logdet} is due to the calculation characteristics of this API.
When the input matrix is close to being non-invertible (\eg, very small singular values), then these APIs may return incorrect results~\cite{torch_doc2}.
Rejected cases show the limitations of \sys in identifying bugs on APIs that are allowed to exhibit anomalous behaviors.
How to identify and filter these unimportant anomalous behaviors to reduce rejected cases will be our future direction.

\noindent
\fbox{\parbox{0.95\linewidth}
{\textbf{Answer to RQ1}:
\sys can effectively utilize the bug reports of source APIs to identify API bugs in their analogous API functions. 
\upd{Response to R3Q1: }{
In the experiment, \sys leverages 172 bug cases to detect 151 API bugs, among which 124 are newly reported and 13 are performance bugs that existing methods fail to detect.
These results demonstrate the effectiveness of \sys in finding a wide range of real-world bugs.
}
Furthermore, 49 of these detected bugs originate from analogous API pairs that existing approaches cannot match, while 9 bugs are from APIs that existing methods do not cover.
These results highlight the contribution of the APIs and API pairs newly covered by \sys in enhancing bug detection.
}}


\subsection{Efficiency in Detecting Bugs}\label{s:effective_rq2}

\noindent
{\bf Experiment Design and Results:}
To evaluate the efficiency of \sys in generating test cases and triggering bugs, we conduct experiments to calculate and compare the \textit{Ratio of test cases that can trigger bugs} and \textit{Average time to detect bugs} of \sys and baseline methods.
Specifically, we execute the complete procedure of \sys and three baselines to record the total number of generated test cases and the number of test cases that can trigger bugs, and the time cost of testing, as described in~\autoref{s:setup}.
As the baselines may not save some test cases or inputs, to ensure that we do not mistakenly count such excluded test cases, we directly record the total number of generated test files as the total number of test cases.
The experimental results are shown in~\autoref{tab:rq1-2}.
The column `\#Valid' shows the number of generated cases that can trigger bugs and the column `\#Total' denotes the overall number of generated cases in one complete execution.
The column `Ratio' displays the ratio of the number of cases that triggered bugs to the total number of generated test cases, corresponding to the `\textit{Ratio of test cases that can trigger bugs}' in~\autoref{s:setup}.
The last column of~\autoref{tab:rq1-2} displays the `\textit{Average time to detect bugs}' of each approach.

\noindent
{\bf Analysis:}
The results in~\autoref{tab:rq1-2} demonstrate the efficiency of \sys in generating test cases to trigger bugs.
In the complete execution of baselines, DocTer generates 33,859 test files on PyTorch and TensorFlow in a complete execution, and only 251 of them (0.74\% of the total) are valid cases.
The testing process on two frameworks lasts over 99 hours, and the average time to detect bugs in DocTer is 41.98 minutes.
On the PyTorch framework, DeepREL spends over 27 hours generating a total of 77,662 test files and marks 2,001 of them as `can trigger bugs', and the valid test case ratio is 2.58\%.
DeepREL does not provide the test cases that can trigger bugs or the corresponding numbers on TensorFlow.
Therefore, we count the number of candidate bugs it detects (the upper bound of possible bug cases) in~\autoref{tab:rq1-2}.
DeepREL spends over 150 hours testing two frameworks, and at most 1.23\% of all 330,195 generated test files can trigger bugs.
The average time to detect bugs of DeepREL is 58.62 minutes.
TitanFuzz spends over 72 hours generating 350,047 cases on two DL frameworks, 3.90\% of which are candidates and can catch valuable buggy behaviors or trigger bugs.
The average time of TitanFuzz to detect bugs is 68.43 minutes.
By contrast, \sys average spends 5.70 minutes to detect one bug, which is only 13.57\%, 9.72\%, and 8.33\% of the average time cost of DocTer, DeepREL, and TitanFuzz, respectively.
\sys is over 10x more time efficient than DeepREL and TitanFuzz in detecting bugs.
In testing, \sys generates a total of 404 test files based on 172 collected bugs within 15 hours, and 143 of them can be used to discover bugs, and the ratio reaches 35.40\%.
Note that quite a part of the generated test files can be used to detect bugs for multiple analogous API functions at the same time, which improves the efficiency of \sys in generating test cases and detecting bugs.
In addition, our experiments show that the time from when \sys starts testing to when it triggers the first bug is within 3 minutes, while baselines usually tend to take 8 minutes or more.
\sys, therefore, outranks three baselines in generating and utilizing test cases to detect bugs efficiently.

\upd{Response to R3Q3: }{
Existing bug-finding tools use various techniques, such as fuzzing, to explore and discover new anomalous behaviors and detect bugs.
As a bug-finding tool orthogonal to existing tools, \sys aims to use reported bugs to discover API bugs on analogous APIs, therefore, it can be used to enhance the effectiveness and efficiency of bug detection.
In the experiment, \sys leverages 13 bug cases from the baseline method reports and successfully detects 18 API bugs within 3 hours, 11 of which are not reported by the baselines.
This demonstrates the potential of collaboration between \sys and existing bug-finding tools to accelerate DL framework bug detection.
}

\noindent
\fbox{\parbox{0.95\linewidth}
{\textbf{Answer to RQ2}:
\sys can efficiently generate test cases to trigger API bugs.
In the experiments, 35.40\% of the test cases generated by \sys can trigger bugs, which is over 9.08 times the bug triggering ratio of the baseline methods.
Furthermore, the average time to detect bugs of \sys is 5.7 minutes, while baseline methods need at least 41.98 minutes, demonstrating the efficiency of \sys in detecting real-world bugs.
}}

\subsection{Impacts of Configurable Parameter}\label{s:configurable}

\begin{figure}
  \centering     
  \includegraphics[width=0.75\textwidth]{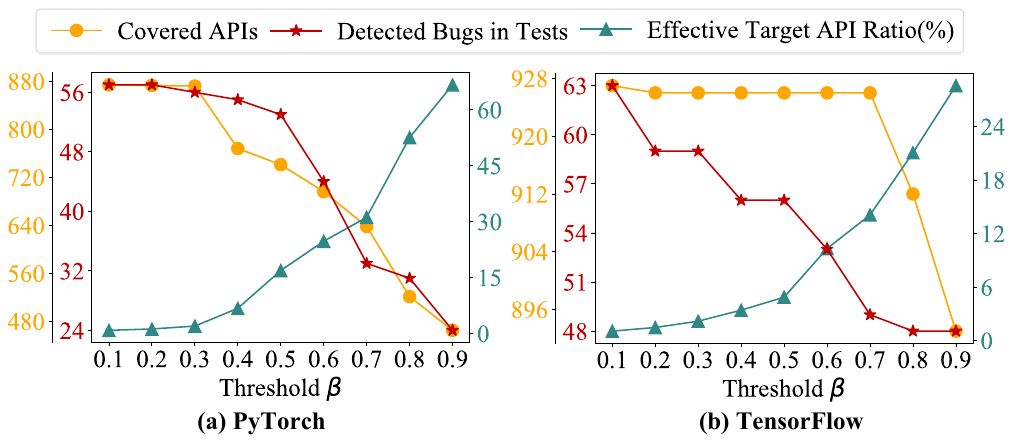}
  \caption{Impacts of Different \(\beta\) Values on PyTorch and TensorFlow}
  \label{fig:rq3}
\end{figure}

\noindent
{\bf Experiment Design and Results:}
\sys leverages the built-in threshold \(\beta\) to determine whether two API functions share context similarity.
To figure out the impacts of the threshold \(\beta\) and select its default value in \sys, we conduct experiments with different \(\beta\) values from 0.1 to 0.9 in increments of 0.1.
\autoref{fig:rq3} show the results of these experiments on PyTorch and TensorFlow respectively.
The X-axis represents the threshold values.
As shown in the legend, the orange and red lines separately indicate the number of API functions covered by context similarity and the number of detected bugs.
The blue line illustrates the ratio of the number of target APIs that successfully trigger bugs to the total number of target APIs matched by context similarity (for convenience, we refer to it as \textit{effective target API ratio}).
A higher effective target API ratio means that API pairs matched by context similarity are more effective in detecting bugs.
Numbers in different colors on the Y-axis show the values of corresponding lines.


\noindent
{\bf Analysis:} 
\autoref{fig:rq3} illustrates the impact of different values of \(\beta\) on the detection results of \sys on different frameworks.
Since modifying \(\beta\) has similar effects on the testing results of different frameworks, we focus our detailed analysis on the impact of \(\beta\) on PyTorch in this section.

When \(\beta\) is set to 0.6 to 0.8, \sys can achieve good detection results.
A higher \(\beta\) value leads to more stringent evaluations of API pairs sharing context similarity, resulting in a reduction in the number of covered API functions, as shown in~\autoref{fig:rq3}(a).
Increasing the threshold value from 0.7 to 0.9 results in a sharp decline in the number of covered API functions from 638 to 465.
Simultaneously, the bug detection ability of \sys also declines significantly.
When \(\beta\) is 0.4, \sys can detect 55 bugs on PyTorch through context-similar API pairs, and this number decreases to only 24 when \(\beta\) increases to 0.9.
However, the increase in the \(\beta\) can improve the effectiveness of matched context-similar API functions in triggering bugs.
When the \(\beta\) increases from 0.4 to 0.9, the effective target API ratio on PyTorch rises from 16.87\% to 66.67\%.
Therefore, there is a trade-off between the ratio of effective target API functions and the number of detected bugs and covered API functions.
Using a higher threshold can retain API pairs that have more similar contexts, but other API pairs that have the potential to trigger bugs will also be discarded, resulting in a degradation of the overall testing effectiveness.
Finally, we set the default value of \(\beta\) to 0.6 on PyTorch and 0.8 on TensorFlow in \sys to maximize its API coverage, effectiveness in bug detection, and effective target API ratio.

\noindent
\fbox{\parbox{0.95\linewidth}
{\textbf{Answer to RQ3}:
The configurable parameter \(\beta\) has a significant impact on the bug detection and API coverage of \sys.
A too large \(\beta\) will discard API pairs that may trigger similar bugs, resulting in a degradation in the effectiveness of bug detection.
\sys selects \(\beta=0.6\) on PyTorch and \(\beta=0.8\) on TensorFlow to maximize the effectiveness of bug detection.
}}

\section{Related Work}
\label{sec:related}

In this paper, we propose \sys that matches analogous DL framework APIs according to context similarity and argument similarity and generates test cases based on real-world bugs.
It is highly related to DL framework testing and code similarity measurement.

\noindent
{\bf DL Framework Testing.}
Researchers have proposed various methods to test DL framework (\eg, PyTorch and TensorFlow) through model-level methods~\cite{gu2022muffin,mu2025improving,wang2024d} and API-level methods~\cite{wang2022eagle,deng2022fuzzing,xie2022docter}, which have been comprehensively introduced in~\autoref{sec:dltest}.
In addition, existing works also design elaborated metamorphic relations to validate the correctness of DL framework implementation~\cite{ding2017validating,wang2020accuracy}.
\upd{Response to R2Q2: }{
FreeFuzz~\cite{wei2022free} successfully detects one performance bug using metamorphic testing techniques.
However, limited by its metamorphic relation, it can only test framework behaviors related to tensor types and cannot effectively identify other diverse performance bugs (e.g., the \texttt{LazyConvTranspose2d} bug in~\autoref{fig:moti} and the \texttt{Hardtanh} bug in~\autoref{fig:rq1-case1}).
}
Recently, Zhang et al.~\cite{zhang2021predoo} propose the test tool `Predoo', which performs a fine-grained evaluation of the shape variable input and error of 7 operators of the DL framework.
Researchers also focus on the security problem of DL frameworks.
SkipFuzz~\cite{kang2022skipfuzz} uses active learning to learn the input constraints of different library API functions and generates valid test inputs for TensorFlow and PyTorch. It had finally identified 43 crashes on DL frameworks, including 13 CVEs assigned.
IvySyn~\cite{christou2023ivysyn} constructs code blocks by DL framework APIs based on a set of offending inputs that trigger memory safety errors in the underlying implementation of DL frameworks (\eg, in C/C++ program language) to trigger security vulnerabilities.
In addition, researchers also detect bugs in other DL underlying libraries such as DL compilers.
Liu et al.~\cite{liu2022coverage} design a testing method for the ML compiler framework TVM.
This method is guided by coverage feedback to mutate the low-level intermediate representation of TVM to achieve more effective fuzzing testing.
Shen et al.~\cite{shen2024tale} transfer the knowledge of DL framework fuzzers (\eg, DocTer and DeepRel) to generate effective test cases for diverse DL compiler operators and detect crashes and inconsistencies.
\upd{Response to R1Q2 and R2Q3: }{
Shiri et al.~\cite{shiri2025history} extract corner case patterns from historical issue reports to guide fuzzers in generating test cases and finding bugs more effectively.
Their method uses differential testing and crashes to construct test oracles and effectively identifies status and value bugs.
However, this design still suffers from the limitation described in~\autoref{sec:moti-new}, namely, the lack of capability of detecting performance bugs.
To effectively test rapidly iterating DL frameworks, Xie et al.~\cite{xie2024cedar} built upon previous work~\cite{xie2022docter,wang2022eagle} to design a continuous testing framework for efficiently discovering regression bugs and masked bugs.
}
Recently, Zhang et al.~\cite{zhang2025deep} survey the testing methods on various DL libraries and point out the challenges of future testing research.

Different from the above methods, \sys is built on the concept of code similarity measurement and real-world bugs confirmed by developers.
\sys focuses on DL framework bugs and can leverage existing bug reports on one API to efficiently exploit bugs on its analogous APIs.

\noindent
{\bf Code Similarity Measurement.}
Existing work has a variety of code similarity measurement and code clone detection methods.
These methods can be divided into static and dynamic according to whether the test code needs to be executed.
The static methods mainly include metrics-based~\cite{ottenstein1976algorithmic,faidhi1987empirical}, text-based~\cite{roy2008nicad,luo2014semantics}, token-based~\cite{sajnani2016sourcerercc,djuric2013source}, AST-based~\cite{zhang2019novel,jiang2007context}, and graph-level methods~\cite{crussell2012attack,chae2013software}.
Among them, the AST and graph-level methods can comprehensively understand the syntax and capture the relationship of the function calls between code blocks, therefore, they have better detection effects but greater overhead than other methods.
Existing dynamic methods propose that the functional similarity between programs can be evaluated from the perspectives of input and output~\cite{elva2012semantic,su2016identifying}, abstract memory state~\cite{kim2011mecc}, etc.
Recently, Maertens et al.~\cite{maertens2022dolos} propose an open-source tool that supports a broad range of programming languages (\eg, C, Java, Python, go).
Zhang et al.~\cite{zhang2022astro} present an AST-assisted approach for generalizable neural clone detection to find clones in codebases reflecting industry practices.
Wang et al.~\cite{wang2023ccstokener} design a code clone detection tool based on the semantic token which enhances the detection capability by complementing the traditional token with semantic information.
Our work mainly combines the concept of code similarity measurement into DL framework testing.
\sys matches context-similar APIs based on the similarity of their context information and leverages the bug reports on one API to efficiently find new API bugs on its analogous APIs.

\section{Discussion}
\label{sec:discuss}

\noindent
\upd{Response to R3Q3: }{\textbf{Discussion.}
\ding{182} \textit{Impact of Collected Bug Cases:}
Different from existing fuzzing methods~\cite{xie2022docter,deng2023large}, which directly generate or mutate test cases for individual APIs, \sys generates new test cases and finds API bugs for analogous APIs based on collected bug cases.
Note that the detection results of \sys are inherently dependent on the collected bug cases.
The more extensive and diverse problematic APIs these cases encompass, the more comprehensive the API coverage \sys can achieve during detection, leading to the discovery of additional previously unknown bugs.
Conversely, a limited number of bug cases concentrated on specific APIs constrain \sys to testing only a narrow subset of analogous APIs.
Although the experiments in~\autoref{s:effective} successfully cover 1,204 DL framework APIs and detect 124 unreported API bugs based on the 172 collected bug cases, demonstrating the effectiveness and efficiency of \sys, this still represents only a small fraction of the matched analogous APIs, as shown in~\autoref{tab:rq1-2}.
Developing techniques to automatically collect and annotate a more diverse set of bug cases, thereby finding bugs on a broader range of APIs, is a promising future research direction.}
\updmj{Response to R2Q2: }{
\ding{183} \textit{Scope of \sys:}
As a testing tool orthogonal to existing bug-finding approaches, rather than searching for unknown API bugs from scratch, \sys leverages a known bug in one API to uncover bugs in its similar APIs.
Note that the API bugs discovered by \sys may stem from the same underlying implementation errors as the known API bugs.
However, this does not diminish the value of \sys.
Effectively identifying potential API bugs is crucial for the development and maintenance of DL frameworks.
For framework users, \sys could promptly recognize APIs exhibiting abnormal behavior, thereby preventing users from misusing these buggy APIs and mitigating potential security risks in users' models and software.
For framework developers, \sys efficiently reveals abnormal behaviors across a series of APIs that share similar underlying implementations, providing developers a more comprehensive debugging perspective, helping them infer the root cause and efficiently localize and fix bugs.
Moreover, the experiment results in~\autoref{s:effective_rq2} demonstrate that \sys utilizes 13 bug cases reported by baselines to detect 18 API bugs within three hours, 11 of which are previously unreported, demonstrating the efficiency and practicality of \sys.
These findings highlight that \sys is not a replacement for existing methods, but rather a complement to existing testing tools.
It can effectively leverage individual bug reports and improve the effectiveness and efficiency of the entire bug detection and repair process.
}

\noindent
\textbf{\sys Enhancement.}
\ding{182} \textit{Automated Verification and Annotation:} Currently, \sys receives the bug case list as one of its inputs, which requires manual intervention.
How to automatically verify and annotate bug cases (\eg, using code models) is a potential future direction to enhance \sys's efficiency.
\ding{183} \textit{Performance Bug Reports:} \sys presently handles performance bug reports that provide code snippets to estimate expected overhead.
A valuable future direction would involve automating the calculation of expected overhead or extracting expected behavior from bug reports described in natural language or images~\cite{torch_perfnlp}.
\upd{Response to R2Q1: }{
\ding{184} \textit{Fault Localization: }
Some API bugs may share a common root cause, and a single effective patch can potentially resolve multiple related issues~\cite{torch_issue3}.
The detection results and corresponding fix patches reported in prior work~\cite{xie2022docter,torch_docterpatch} further support this observation.
Note that \sys, like existing bug-finding tools~\cite{xie2022docter,deng2022fuzzing}, is designed for bug detection rather than faulty localization or program repair.
As such, identifying the root causes of API bugs lies outside the scope of this testing work.
We have reported the number of duplicate bugs and the number of API bugs that share the same implementation errors as the source bugs in~\autoref{s:effective}.
However, since our reports for TensorFlow API bugs currently do not receive any patches, we are unable to conduct the analysis on this framework.
We estimate that API bugs with shared underlying implementations also exist in this framework.
Therefore, designing error localization methods for automated fault localization remains a valuable direction for future research, which could effectively reduce the debugging workload for DL framework developers (e.g., TensorFlow) and improve bug repair efficiency.
}
\ding{185} \textit{Code Similarity Measurement:}
To mitigate false positives in the static analyzer, \sys selects strict thresholds to judge analogous source code functions based on the experimental results of prior work~\cite{su2016identifying,misu2017interface}.
Even if there are still a few false positives in the static analyzer, they can hardly affect the subsequent API matching and bug detection.
On the one hand, API functions that are mistakenly matched due to these false positives are easily discarded by the API matcher due to argument mismatches.
On the other hand, \sys identifies bugs when the target API exhibits the same buggy behavior that is consistent with the source API's bug report.
Therefore, \sys can always detect real anomalous behavior.
However, the static analyzer could still miss some analogous source code functions.
Specifically, different developers may use the same logic to construct functions to solve specific problems, resulting in different implementations of two source code functions but similar functionality~\cite{zakeri2023systematic}.
Currently, the static analyzer cannot effectively identify such analogous functions.
How to enhance the matching effect of similar source code functions and APIs in \sys will be a future direction.
\upd{Response to R1Q3 and R3Q3: }{
\ding{186} \textit{Integration with Existing Tools:}
The detection effectiveness of \sys is influenced by the diversity and quality of the collected bug cases.
Therefore, as described in~\autoref{sec:sampler}, we collect and validate bug cases in the preparation stage to ensure the quality of the bug cases extracted from GitHub repositories.
A promising future direction is to integrate \sys with other bug-finding tools (e.g., fuzzing) into a unified testing pipeline.
In such a pipeline, once an API bug is detected by an external tool, the problematic API and its bug-triggering test case can be directly used as input to \sys.
\sys can then efficiently construct new test cases and detect bugs in analogous APIs.
This workflow has the potential to substantially reduce manual effort, enhance the efficiency of bug detection, and assist developers in more effectively identifying the root causes of bugs.
}

\noindent
\textbf{DL Bug Finding.}
\ding{182} \textit{LLM Library Bug Finding:} 
With the advent of Large Language Model (LLM) technology, dependent libraries (\eg, APEX~\cite{apex}) of LLMs exhibit various bugs, such as crashes during model training~\cite{zhang2022opt,mu2025designing,jiang2025foundation}, which impede the application and deployment of LLMs.
However, the substantial runtime overhead of LLMs renders traditional DL fuzz testing methods, which generate millions of test cases, unsuitable for LLM library testing.
Developing an efficient LLM library testing method to uncover potential bugs will be a valuable future direction.
\ding{183} \textit{Performance Bug Detection:} 
Current DL framework testing primarily addresses crashes and numerical inconsistencies, with little attention given to performance bugs that impact model training, economy, and even the environment.
Detecting performance bugs necessitates constructing a test oracle and estimating the expected overhead.
Our finding suggests that different settings of API arguments may influence the actual overhead (\eg, `Bug Case 1' of~\autoref{s:effective}).
Therefore, the runtime overhead of one API function can be qualitatively or even quantitatively estimated based on the description of these arguments, serving as a pseudo-test oracle.
Formalizing expected performance changes from API argument descriptions and constructing test oracles to detect performance bugs will be an important future direction.


\section{Conclusion}
\label{sec:conclusion}

This paper presents \sys, a bug-finding tool for DL frameworks that can find new bugs that are similar to known bugs, regardless of bug types.
For a problematic DL framework API function and its associated bugs, it matches analogous API functions from the perspectives of context similarity and signature similarity, and then synthesizes test cases for these analogous API functions.
Then, \sys leverages the behavior of the confirmed bug on the problematic API as the test oracle to evaluate the generated test cases and efficiently identify new API bugs on analogous API functions.
Our evaluation on two frameworks shows that \sys can effectively and efficiently detect status, value, and performance bugs.

\begin{acks} 
The authors would like to thank the anonymous reviewers for their insightful comments and valuable suggestions.
Authors in China are supported partially by the National Key Research and Development Program of China (2023YFE0209800), the National Natural Science Foundation of China (U24B20185, T2442014, 62161160337, 62132011, U21B2018), and the Shaanxi Province Key Industry Innovation Program (2023-ZDLGY-38, 2021ZDLGY01-02).
Thanks to the New Cornerstone Science Foundation and the Xplorer Prize.
\end{acks}
\newpage

\bibliographystyle{ACM-Reference-Format}
\bibliography{software}

\end{document}